\documentclass[twoside,twocolumn,9pt]{article}
\usepackage{extsizes}
\usepackage[super,sort&compress,comma]{natbib}
\usepackage[version=3]{mhchem}
\usepackage[left=1.5cm, right=1.5cm, top=1.785cm, bottom=2.0cm]{geometry}
\usepackage{balance}
\usepackage{mathptmx}
\usepackage{sectsty}
\usepackage{graphicx}
\usepackage{lastpage}
\usepackage[format=plain,justification=justified,singlelinecheck=false,font={stretch=1.125,small,sf},labelfont=bf,labelsep=space]{caption}
\usepackage{float}
\usepackage{fancyhdr}
\usepackage{fnpos}
\usepackage{algorithmic}
\usepackage{algorithm}
\usepackage{amsmath,amssymb,amsfonts,textcomp}
\usepackage{float,flafter}
\usepackage[utf8]{inputenc}
\usepackage{pgfplots}
\pgfplotsset{compat=1.9}
\usepackage{filecontents}
\usepackage{inputenc}
\usepackage[english]{babel}
\addto{\captionsenglish}{%
	
}
\usepackage{array}
\usepackage{droidsans}
\usepackage{charter}
\usepackage[T1]{fontenc}
\usepackage{setspace}
\usepackage[compact]{titlesec}
\usepackage{hyperref}
\usepackage{xurl}
\usepackage{epstopdf}
\newcommand{\commentOut}[1]{}
\definecolor{cream}{RGB}{222,217,201}
\definecolor{purple1}{rgb}{0.5,0,0.75}

\begin{document}
\pagestyle{fancy}
\thispagestyle{plain}
\fancypagestyle{plain}{
	\renewcommand{\headrulewidth}{0pt}
}

\makeFNbottom
\makeatletter
\renewcommand\LARGE{\@setfontsize\LARGE{15pt}{17}}
\renewcommand\Large{\@setfontsize\Large{12pt}{14}}
\renewcommand\large{\@setfontsize\large{10pt}{12}}
\renewcommand\footnotesize{\@setfontsize\footnotesize{7pt}{10}}
\renewcommand\scriptsize{\@setfontsize\scriptsize{7pt}{7}}
\makeatother

\renewcommand{\thefootnote}{\fnsymbol{footnote}}
\renewcommand\footnoterule{\vspace*{1pt}%
	\color{cream}\hrule width 3.5in height 0.4pt \color{black} \vspace*{5pt}}
\setcounter{secnumdepth}{5}

\makeatletter
\renewcommand\@biblabel[1]{#1}
\renewcommand\@makefntext[1]%
{\noindent\makebox[0pt][r]{\@thefnmark\,}#1}
\makeatother
\renewcommand{\figurename}{\small{Fig.}~}
\sectionfont{\sffamily\Large}
\subsectionfont{\normalsize}
\subsubsectionfont{\bf}
\setstretch{1.125} 
\setlength{\skip\footins}{0.8cm}
\setlength{\footnotesep}{0.25cm}
\setlength{\jot}{10pt}
\titlespacing*{\section}{0pt}{4pt}{4pt}
\titlespacing*{\subsection}{0pt}{15pt}{1pt}
\newcolumntype{L}{>{\centering\arraybackslash}m{3.5cm}}

\fancyfoot{}
\fancyfoot[LO,RE]{\vspace{-7.1pt}\includegraphics[height=9pt]{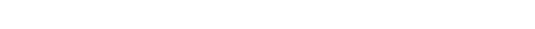}}
\fancyfoot[CO]{\vspace{-7.1pt}\hspace{13.2cm}\includegraphics{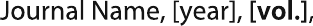}}
\fancyfoot[CE]{\vspace{-7.2pt}\hspace{-14.2cm}\includegraphics{head_foot/RF}}
\fancyfoot[RO]{\footnotesize{\sffamily{1--\pageref{LastPage} ~\textbar  \hspace{2pt}\thepage}}}
\fancyfoot[LE]{\footnotesize{\sffamily{\thepage~\textbar\hspace{3.45cm} 1--\pageref{LastPage}}}}
\fancyhead{}
\renewcommand{\headrulewidth}{0pt}
\renewcommand{\footrulewidth}{0pt}
\setlength{\arrayrulewidth}{1pt}
\setlength{\columnsep}{6.5mm}
\setlength\bibsep{1pt}

\makeatletter
\newlength{\figrulesep}
\setlength{\figrulesep}{0.5\textfloatsep}

\newcommand{\topfigrule}{\vspace*{-1pt}%
	\noindent{\color{cream}\rule[-\figrulesep]{\columnwidth}{1.5pt}} }

\newcommand{\botfigrule}{\vspace*{-2pt}%
	\noindent{\color{cream}\rule[\figrulesep]{\columnwidth}{1.5pt}} }

\newcommand{\dblfigrule}{\vspace*{-1pt}%
	\noindent{\color{cream}\rule[-\figrulesep]{\textwidth}{1.5pt}} }

\makeatother

\twocolumn[
	\begin{@twocolumnfalse}
		{\includegraphics[height=30pt]{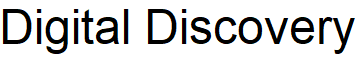}\hfill\raisebox{0pt}[0pt][0pt]{\includegraphics[height=55pt]{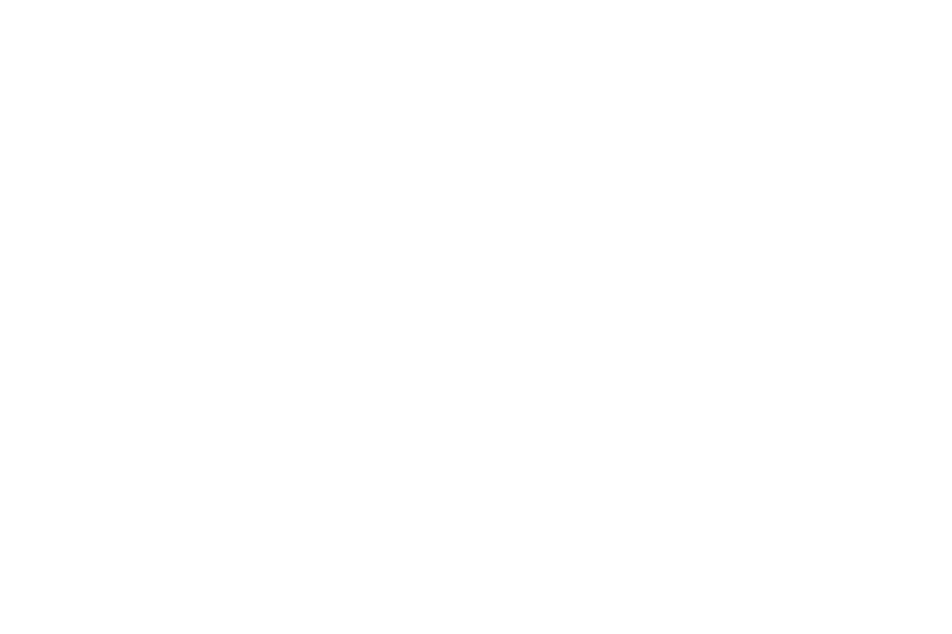}}\\[1ex]
			\includegraphics[width=18.5cm]{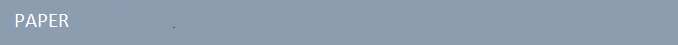}}\par
		\vspace{1em}
		\sffamily
		\begin{tabular}{m{4.5cm} p{13.5cm} }

			\includegraphics{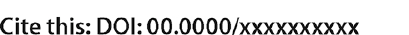}   & \noindent\LARGE{\textbf{Automated Routing of Droplets for DNA Storage on a Digital Microfluidics Platform}}                                     \\
			                                  & \vspace{0.3cm}                                                                                                                                  \\

			                                  & \noindent\large{Ajay Manicka, Andrew Stephan, Sriram Chari, Gemma Mendonsa, Peyton Okubo, John Stolzberg-Schray, Anil Reddy, and Marc Riedel} \\

			\includegraphics{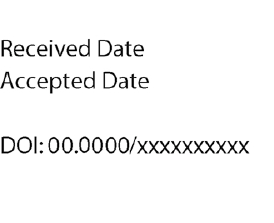} &                                                                                                                                                 \\
		\end{tabular}

	\end{@twocolumnfalse} \vspace{0.6cm}

]

\renewcommand*\rmdefault{bch}\normalfont\upshape
\rmfamily
\section*{}
\vspace{-1cm}







\sffamily{\textbf{Technologies for sequencing (reading) and synthesizing (writing) DNA have progressed on a Moore's law-like trajectory over the last three decades. This has motivated the idea of using DNA for data storage. Theoretically, DNA-based storage systems could out-compete all existing forms of archival storage. However, a large gap exists between what is theoretically possible in terms of read and write speeds and what has been practically demonstrated with DNA. This paper introduces a novel approach to DNA storage, with automated assembly on a digital microfluidic biochip. This technology offers unprecedented parallelism in DNA assembly using a dual library of ``symbols'' and ``linkers''. An algorithmic solution is discussed for the problem of managing droplet traffic on the device, with prioritized three-dimensional ``A*'' routing. An overview is given of the software that was developed for routing a large number of droplets in parallel on the device, minimizing congestion and maximizing throughput.}}


\rmfamily 


\section{Introduction} \label{sec:intro}

\subsection{The World of Information}

The amount of data that the world generates has been increasing exponentially since the inception of the computer age. This trend will continue for the foreseeable future, as ever more IoT devices come online and humans create denser content in the form of video and virtual reality. The bulk of this so-called ``big data'' is stored in hard disk drives (HDDs)~\cite{li2017hard}. It is estimated that the total demand for data storage by 2025 will be 180 zettabytes (1 zettabyte = 1 billion terabytes)~\cite{reinsel2021worldwide}, which would be a three-fold increase from 2020. Some of this newly generated data will need to be archived for long-term storage, perhaps 9.3ZB of it~\cite{archiving}. Even if only 5\% of stored data is placed in ``deep'' offline archives, this would require 46.5 million 20TB HDDs by 2025. The required capacity for online and on-premise archiving will be many times larger. 

Meanwhile, the supply of storage media is projected to grow by less than 20\% year over year in the same time-frame~\cite{IDC2021storage}. Without the construction of new HDD and SSD manufacturing facilities, which are multi-billion dollar investments, demand for storage is expected to outstrip supply by as much as two-fold~\cite{gartnerMarketTrends}. Furthermore, magnetic storage has durability limitations that make it undesirable for maintenance-free, multi-decade storage. Also, the proliferation of data centers is causing long-term environmental damage, as the electricity they require, mostly for cooling, is a major source of global carbon emissions\cite{Leproust2022Data}. For all these reasons, there has been a strong interest in identifying new types of storage media.

A strong contender for a type of media that could meet the future demand for archival storage is DNA. The theoretical storage capacity of DNA is as high as 200 petabytes per gram, which is over a thousand times denser than conventional HDDs~\cite{church12, seagate.com}. Most importantly, the energy requirement for writing is on the order of $10^{-19}$ joules per bit which is orders of magnitude below the femtojoules/bit ($10^{-15}$ joules per bit) barrier touted for other emerging technologies~\cite{church12}. The durability of DNA is unmatched, exceeding centuries, while hard drives and magnetic tape rarely maintain reliability longer than 30 years \cite{el2022high}. We point to a review paper that summarizes the potential of DNA storage systems~\cite{ceze19}.

\subsection{DNA as a Storage Unit}

\begin{figure}[h]
	\centering
	\vspace{0.5cm}
	\includegraphics[scale=0.6]{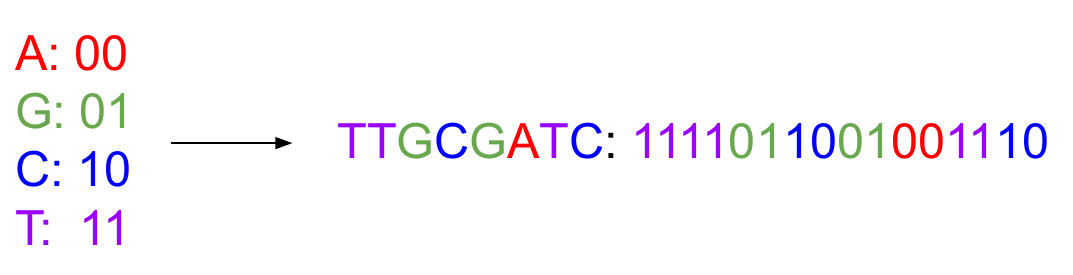}
	\caption{This figure represents the mapping of nucleotide bases to a binary code. Just as we can string together binary values, we can assemble DNA nucleotides chemically to represent data.}
	\label{fig:BaseToBinary}
\end{figure}

Traditional computer systems use the binary code of zeros and ones $\{0,1\}$ as storage units. In its simplest form, a DNA storage system uses a quaternary code of nucleotides drawn from four different nitrogenous bases, viz. adenine, guanine, cytosine, and thymine, denoted $A, G, C$, and $T$, respectively. We can map couplets of zeros or ones directly to each nucleotide, as illustrated in~Fig.~\ref{fig:BaseToBinary}. In this way, we use a string of nucleotides to represent arbitrary data. 

With such an encoding, a DNA sequence with $n$ nucleotides stores $2n$ bits of data based on binary mapping. Table \ref{symbol_table} introduces our concept of a DNA Symbol Library.\footnote[1]{A \textbf{DNA Symbol Library} is a set of nucleotide sequences, of some fixed length $n$, which we can use as building blocks to assemble larger DNA storage units.} (We note here that we use footnotes to define terms throughout the text.) From a base set of the 4 nucleotides $\{A, G, C, T\}$, there are 16 ways to select base pairs of length 2; these base pair symbols correspond to binary numbers from 0000 to 1111. If a 2-nucleotide symbol can represent 16 distinct binary numbers, then a 3-nucleotide symbol can represent 64 binary distinct numbers. In general, the addition of each nucleotide quadruples the range of numbers we can represent. So $n$ base pairs can represent $4^n$ distinct numbers for $n > 0$.

\begin{table}[h]
	\centering
	\caption{DNA Symbol Library size based on Symbol Length.}
	\begin{tabular}{|L|L|}
		\hline
      Symbol Length                     & DNA Symbol Library Size                                        \\ 
		(Number of base pairs per symbol) & (Number of Unique Symbols)                         \\
		\hline\hline
		1                                 & 4                                                              \\
		\hline
		2                                 & 16                                                             \\
		\hline
		3                                 & 64                                                             \\
		\hline
		4                                 & 256                                                            \\
		\hline
		5                                 & 1024                                                           \\
		\hline
		6                                 & 4096                                                           \\
		\hline
		7                                 & 16384                                                          \\
		\hline
		8                                 & 65536                                                          \\
		\hline
	\end{tabular}

	\label{symbol_table}
\end{table}

Ever since Watson and Crick first described the molecular double helix structure of DNA~\cite{watson1953structure}, its potential for storage has been apparent to computer scientists. It seems that most practical work is based on liquid-handling robotics. The power consumption of liquid-handling DNA storage systems is on the order of \emph{hundreds} of joules/sec~\cite{ceze19} for a DNA synthesis rate on the order of kilobytes/sec. Overall, these machines use a substantial amount of energy for limited gain. Many creative ideas and novel technologies, ranging from nanopores~\cite{chen20} to DNA origami~\cite{dickinson21}, are also being investigated. The leading approach appears to be phosphoramidite chemistry~\cite{meares2022synthesis}. 


The main barrier to building DNA storage systems that can compete with existing forms of archival storage is the \textit{write} speed, so the rate of DNA synthesis. Hard drive write speeds hover around 50 to 120 MB/s~\cite{hepisuthar2021comparative} while solid-state storage systems achieve write speeds exceeding 200 Megabytes per second~\cite{hepisuthar2021comparative}. All existing DNA storage systems have write speeds many magnitudes slower than this~\cite{erlich2017dna}. 

This paper does not consider the process of reading data stored in DNA (i.e., sequencing it). With current technology, reading DNA is orders of magnitude more efficient than writing it, so the impetus is improvements in write speed. Of course, a complete solution must consider both operations. Nanopore-based devices for sequencing DNA could provide the requisite technology~\cite{jain2016oxford}, as they are compatible with the digital microfluidic technology discussed here.

\subsection{A Solution: Increasing Rate of DNA Synthesis}

Achieving practically useful write speeds will require two things. First, a way to introduce massive parallelization. Second, a chemical protocol that writes as much data as possible per operation, thereby increasing the bit rate (write speed in bits per unit time). This paper proposes a solution to the synthesis speed problem with a dual library of ``symbols\footnote[2]{A \textbf{symbol} is a short double-stranded sequence of DNA whose nucleotides specify the data that is being stored.}'' and ``linkers\footnote[3]{A \textbf{linker} is a double-stranded nucleotide sequence that connects two symbols together in the correct order.}''. The two libraries work in tandem to allow the synthesis of long ``genes\footnote[4]{A \textbf{gene} is a unit of storage consisting of a sequence of symbols, joined by linkers. It is the full length of data assembled as a single molecule of DNA. To be clear, we use the term ``gene'' but the DNA here has no biochemical function; it is only used for storage.}'' each with symbols in the required order, corresponding to the data that is being written. With linkers attached, multiple symbols can be attached to one another in the same droplet. Accordingly, massive parallelization is possible. This is in contrast to most existing schemes for DNA storage, in which each operation attaches a single nucleotide to the end of the sequence, for instance with phosphoramidite chemistry~\cite{twistbioscience}. Details regarding our scheme with symbols and linkers are given in Section~\ref{sec:DNA Assembly}.

Instead of liquid-handling robots, we 
perform assembly of DNA with a digital microfluidic biochip (DMFB). This technology offers the advantages of low reagent consumption, high precision, and miniaturization~\cite{guo2022survey}. Further details are given in Section~\ref{sec:dmf}. 

\begin{figure*}[h]
	\centering
	\includegraphics[scale=0.36]{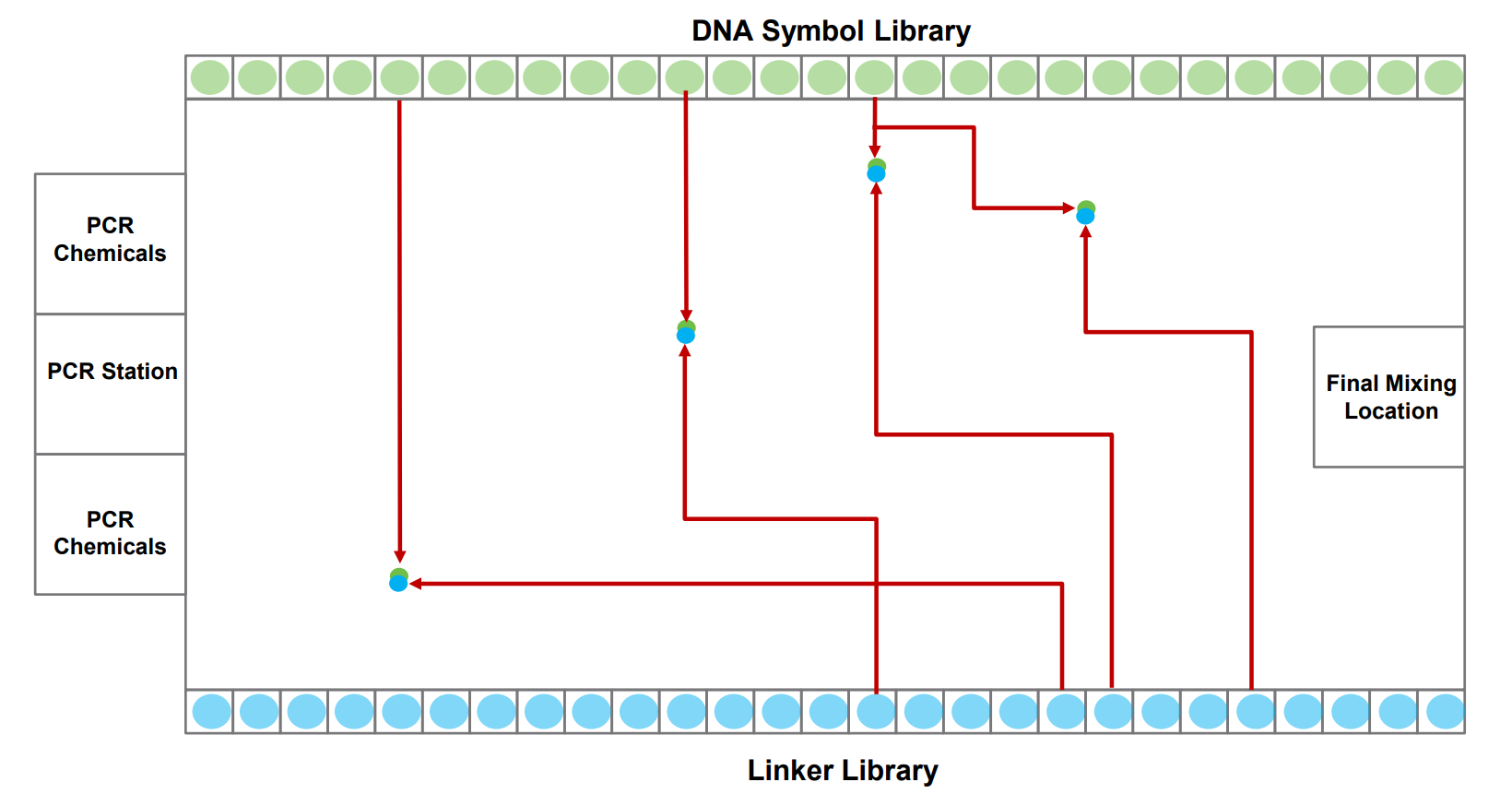}
	\caption{A high-level visual of a digital microfluidic biochip (DMFB). Two sides contain the DNA oligos corresponding to the symbols and linkers. The other two sides contain Gibson, PCR, and ``Purify'' sites, along with the chemicals necessary to complete each chemical process. The figure illustrates the path that eight droplets take, each containing a symbol or a linker. Each symbol meets up with a linker, merging into a larger droplet. The resulting droplet will then be routed to a Gibson site, where the symbol and linker are chemically joined.}
	\label{fig:DMF_Example}
\end{figure*}

A DMFB device can be idealized as a 2-D grid, shown in Fig.~\ref{fig:DMF_Example}. Most of the 2-D grid serves to route individual droplets. In our device, a subset of the available grid points performs dedicated operations: Gibson assembly (\textit{concatenation})~\cite{de2020rapid}; polymerase chain reaction, or PCR (\textit{replication})~\cite{mackay2002real}; and purification (\textit{correction}). One edge of the biochip houses short fragments of DNA in the form of the symbols while the opposite edge holds short fragments in the form of linkers. Also, one of the edges houses PCR stations where depleted stores of DNA symbols and linkers can be refilled. Gibson sites -- locations where symbols are linked together -- and purification sites are strategically positioned throughout the device. The Gibson assembly process is discussed in more detail in Section~\ref{sec:gibson}. We do not discuss the purification process in this paper.

The task of writing DNA begins with an encoding of the data in a gene. When the order to assemble a certain gene is received by the device, it dispenses the requisite DNA symbols, linkers, and chemical reagents as individual droplets along the grid's edge. The droplets corresponding to symbols and linkers contain oligos\footnote{An \textbf{oligo} is a relatively short fragment of DNA.}. When a symbol droplet\footnote{A \textbf{symbol droplet} is a droplet containing a symbol.} and a linker droplet\footnote{A \textbf{linker droplet} is a droplet containing a linker.} meet at a grid point, they merge forming a larger droplet. This larger droplet is routed so that it meets and merges with a chemical droplet\footnote{A \textbf{chemical droplet} is a droplet containing the necessary enzymes for Gibson assembly}. This larger droplet is then routed to a station for Gibson assembly: this is where the symbol and linker oligos are chemically joined to form a single DNA strand.

Routing all the droplets is a significant challenge, one that we confront in this paper. The routing problem becomes more complex as more droplets are pulled to assemble longer genes. 
To solve the problem, we use an algorithm called 3-D prioritized A*\footnote[6]{\textbf{3-D prioritized A*} was selected as the routing algorithm of choice because it is a complete~\cite{firmansyah2016comparative} and optimal~\cite{yao2010path} heuristic-based algorithm that is guaranteed to find the shortest route between a start and goal point, even in the presence of obstacles.~\cite{martins2022improved}}. The algorithm considers three dimensions: the horizontal axis of the DMFB grid, the vertical axis of the DMFB grid, and the axis of time. It is called \textit{prioritized} because it chooses to create routes for droplets by giving priority to the droplet which is furthest from its goal node, i.e., the droplet which has to travel the largest distance across the grid to reach its intended target location. The routing algorithm allows many droplets to move simultaneously while avoiding unwanted collisions. It also allows individual droplets to take the optimal path within the constraints given to them by higher priority droplets. Further details are given in Section~\ref{sec:A_star}.

\subsection{Related Work}\label{sec:related}

DNA storage technology is a rapidly expanding area of research. Here, we reference some relevant literature~\cite{church12, goldman2013towards, grass2015robust, bornholt2016dna, blawat2016forward, erlich2017dna} in the DNA storage space. Our approach to DNA storage differs from prior work: 1) in our use of DMFB technology; and 2) with our novel ``symbol'' and ``linker'' dual library.

With respect to the routing algorithm that we use, this paper builds upon an extensive body of prior work. Numerous papers have discussed routing on DMFB devices, for a variety of applications~\cite{su2006droplet, xu2007integrated, zhao2012simultaneous}. Many discussed versions of the A* approach that we use~\cite{bohringer2004towards, su2006droplet, tsung2009fast}. Other strategies have been considered, for instance, using an evolutionary multi-objective optimization algorithm~\cite{juarez2018evolutionary}. 
Many papers discuss exciting applications of DMFB, for instance, DNA sequencing and clinical diagnosis~\cite{liu2013sample, lehotay2015sampling, perut2016cell}.

Almost all prior work has considered routing on small grids, generally less than 50 $\times$ 50 in size. DNA storage presents very different constraints, particularly with respect to the size of the grid and the degree of parallelism. We target grid sizes that are 6 to 7 orders of magnitude larger. For such large grid sizes, algorithmic runtime is the preeminent concern.

\subsection{Organization}\label{sec:org}

The contents of this paper are organized as follows. First, in Section~\ref{sec:background}, we provide some  background information on DMFB technology and DNA synthesis. Then, in Section~\ref{sec:writespeed}, we discuss the target write speeds and the parameters of a DMFB device that can achieve them. Next, in Section~\ref{sec:DNA Assembly}, we describe our method of automating DNA assembly, with symbols and linkers. Then we discuss the routing requirements for this scheme. Next, in Section~\ref{sec:Architecture}, we describe the software architecture of our system. Then, in Section~\ref{sec:Results}, we present simulation results characterizing the performance, runtime, and memory usage of our software. Finally, in Section~\ref{sec:Conclusions}, we summarize the main results of the paper and discuss areas for future work.


\section{Background}\label{sec:background}

\subsection{Digital Microfluidics (DMF) Technology} \label{sec:dmf}

\begin{figure}[h]
	\centering
	\includegraphics[scale=1.5]{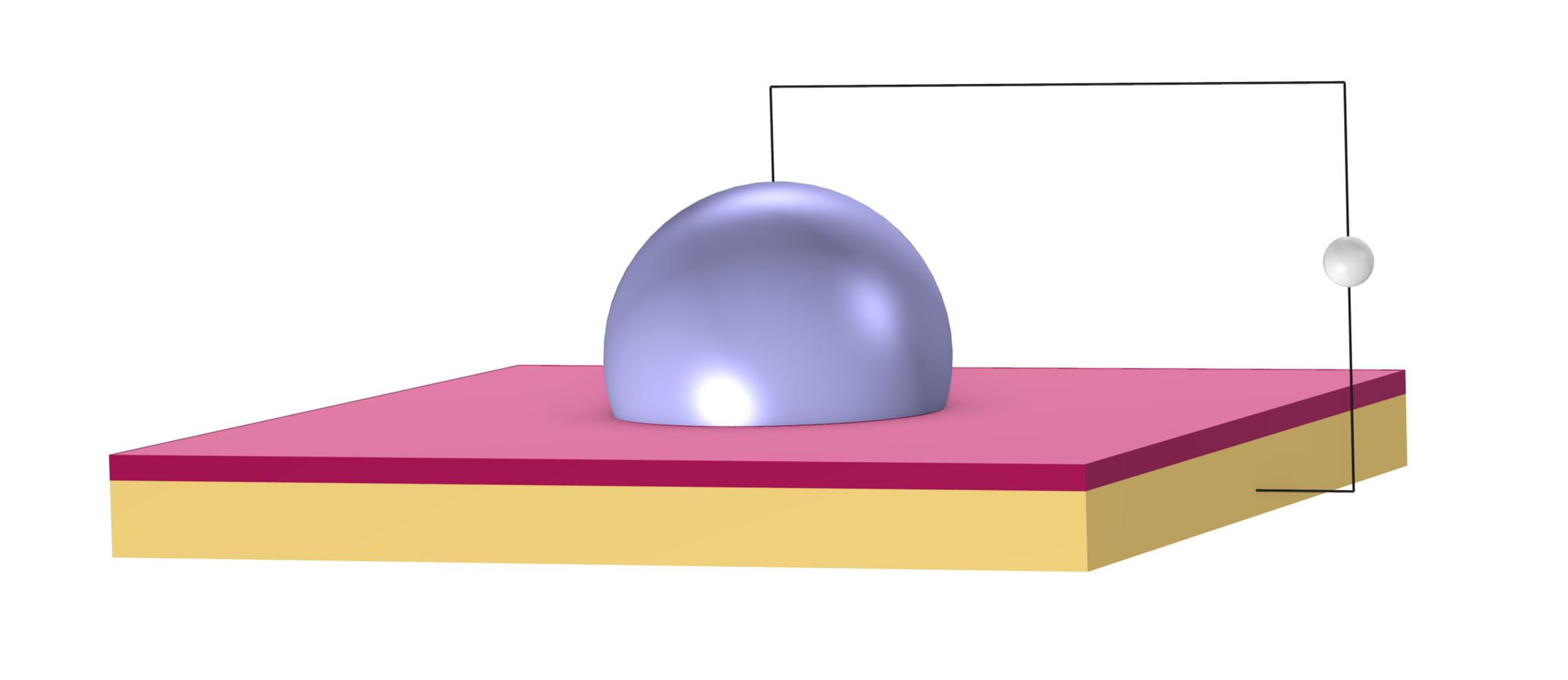}
	\includegraphics[scale=1.5]{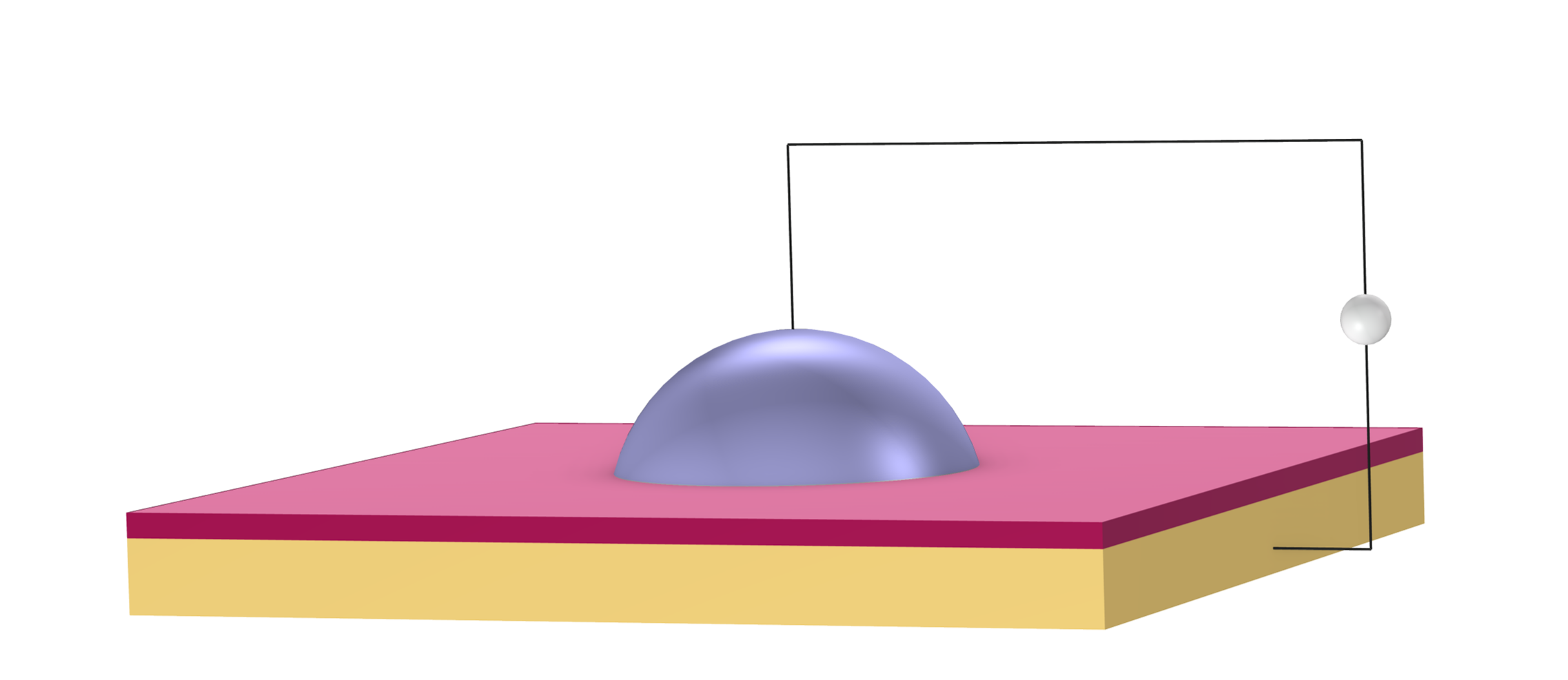}
	\caption{Electrowetting: aqueous droplets spread when a voltage is applied between a droplet and an insulating electrode (shown in orange). The droplet rests upon a hydrophobic dielectric (shown in red). Top: no voltage. Bottom: high voltage.}
	\label{fig:electrowetting}
\end{figure}

DMF is a fluid-handling technology that precisely moves small droplets on a grid by manipulating electrical charge. It works on the principle of \emph{electrowetting}~\cite{mugele2005electrowetting}.  Aqueous droplets naturally bead up on a hydrophobic surface. However, a voltage applied between a droplet and an insulated electrode causes the droplet to spread out on the surface, as shown in Fig.~\ref{fig:electrowetting}.


Electrical signals are applied to an array of such electrodes. Droplets are moved by turning the voltage on and off in succession across adjacent electrodes. The same mechanism can be used to dispense, merge, and mix droplets. These basic operations become the building blocks to perform biochemical reactions. DMFB technology reduces the volume of fluid, and so generally reduces  the cost, compared to technology like liquid-handling robotics~\cite{bender2016digital}. 
It has been studied extensively in academia~\cite{fair07}, and in recent years, has been applied for specific tasks in industry~\cite{millington18}. However, it is fair to say that DMFB remains a niche technology. Scaling down the size of the droplets and increasing the grid dimensions, so increasing the number of droplets on the device, is an expensive proposition in terms of research and development~\cite{yang21}. 

We are working with proprietary DMFB technology that Seagate, a leading storage technology company, is developing. It is of a much greater scale than has been previously demonstrated with very large grid sizes -- millions to billions of electrodes. The technology is not the focus of this paper. Nevertheless, the concepts that we present for DNA storage are predicated on it. In particular, we formulate algorithms to tackle the routing of large numbers of droplets in parallel across Seagate's DMFB platform. Achieving high data throughout, in terms of DNA storage units synthesized per unit of time, is the main objective. 

\subsection{Gibson Assembly Protocol} \label{sec:gibson}

The synthesis of data in the form of DNA begins with DNA fragments, using a process called ``Gibson Assembly''. In 2009, Gibson et al. proposed a method for joining multiple DNA fragments in a single reaction\cite{gibson2009enzymatic}. 
These fragments must have overlapping ends, several base pairs in length. In addition to the fragments to be assembled, three enzymes are also required: exonuclease, DNA polymerase, and DNA ligase. Using these enzymes, Gibson assembly connects two double-stranded DNA together. 

\begin{figure}[h]
	\centering
	\includegraphics[scale=0.4]{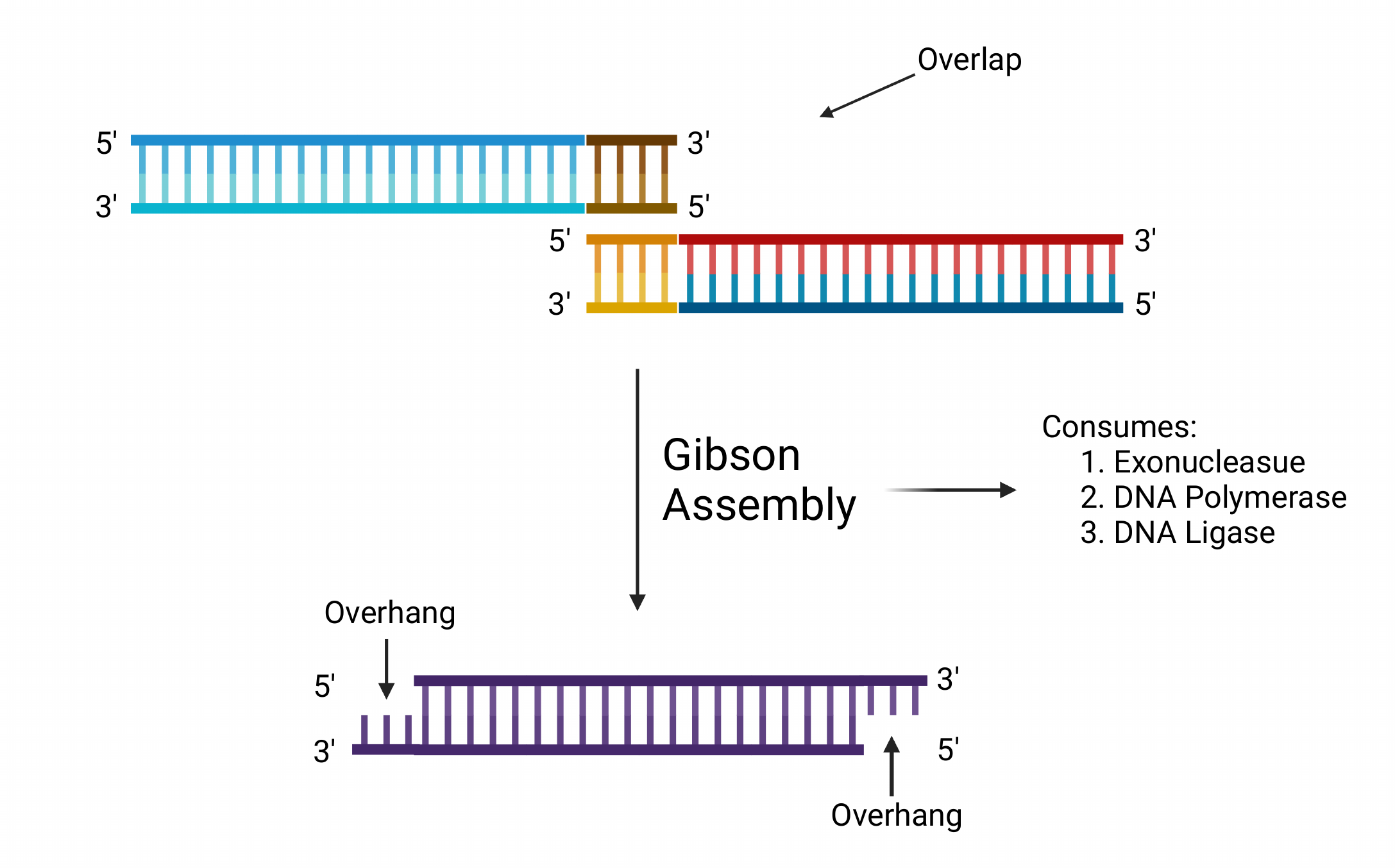}
   \caption{A high-level illustration of the Gibson assembly protocol. It begins with two separate overlapping DNA strands and ends with a single combined strand with two overhangs. Attribution: this figure was created with the software package~\emph{Biorender}.}
	\label{fig:Gibson_assembly_overview}
	
\end{figure}

The Gibson assembly is general-purpose and widely used for cloning DNA fragments. We adapted it to constructing long data-storage strands. A visualization of the Gibson assembly process is shown in Fig. \ref{fig:Gibson_assembly_overview}.

In the context of the DMFB, we place a symbol, linker, and three enzymes into three separate droplets. These are all routed to a ``Gibson Site''. The chemical reactions for Gibson assembly are performed at this site, resulting in a larger droplet with the symbol and linker combined.

\section{Write Speeds}\label{sec:writespeed}


Here, we discuss the parameters of a DMFB that could compete with hard disk drives (HDDs) in terms of data write speed. The write speed of modern HDDs is on the order of 100 megabytes\footnote{A megabyte is $10^6$ bytes. A byte is 8 bits of data.} per second (MB/s), so this is our target write speed:\vspace{0.25cm}\\
\vspace{0.25cm}\noindent\mbox{} \centerline{\bf Goal: A DMF system that writes 100MB of data per second.} \\
We need to load 100MB of data in the form of DNA symbols onto the device per second. Each symbol is 8 nucleotides long, with 2 bits per nucleotide, so each symbol represents 16 bits or 2 bytes of information. Therefore, to meet our target write speed, we must load 50 million symbols, each in a separate droplet, onto the DMFB every second. Using milliseconds as our unit of time, we must load 50,000 symbols per millisecond. We must also load linkers and different chemical reagents. Loading and moving so many droplets is a demanding task -- one that requires mature DMF technology. To achieve our target write speed, we estimate that the DMF device must have the following parameters:

\medskip
\begin{enumerate}
\item A grid size of 100,000 by 100,000 electrodes -- so 10 billion electrodes.

\item Droplet reservoirs or sinks at all four edges -- so 400,000 total. \footnote{Droplets are loaded onto the device from \emph{reservoirs} and loaded off the device into \emph{sinks}.}

\item The ability to load droplets onto the device at a rate of 5kHz -- so a new droplet loaded every 1/5,000-th of a second, or \(200\ \mu \)s. Also, the ability to move droplets from electrode to electrode across the device at the same rate.

\item Handling of droplets that are on the order of a femtoliter in volume.

\end{enumerate}

With four edges to a square grid, symbol droplets will be loaded onto the device from the first edge; linker droplets from the second edge; and various reagents from the third. Completed genes will be loaded off the devices from the fourth edge.

Given the target of loading 50,000 symbol droplets per millisecond onto the chip, with a loading rate of 5kHz we must load 10,000 symbol droplets simultaneously every \(200\ \mu \)s. (We assume that we must also load 10,000 linker droplets and 10,000 chemical droplets simultaneously every \(200\ \mu \)s.) With 100,000 reservoirs containing symbols arranged along the first edge of the device, we assume that a symbol is loaded from 1 out 10 reservoirs every \(200\ \mu \)s.

DMF technology that moves droplets at this speed has been demonstrated~\cite{basova2015droplet,shim2013ultrarapid,li2021active}. However, no DMFB with millions, let alone billions, of electrodes has been built. Undoubtedly, building such a large DMFB is an expensive proposition. 
We note that there is a trade off between DMFB size and droplet speed: the required size of the DMFB decreases as the speed of droplets increases. 
If speed of DMF technology improves beyond 5 kHz, the required grid size will decrease.

The write speed is measured by the number of completed storage genes produced per second (given a gene length of fixed size). We make the following assumptions:

\begin{enumerate}
    \item Each storage gene is assembled from 10,000 symbols.

    \item The system can assemble 10,000 storage genes concurrently.

    \item Assembly is pipelined.\footnote{\textbf{Pipelining} in this context means that we do not wait until assembly of a storage gene is complete before beginning assemble of the next. We begin assembly of the next immediately only one time step later. As a result, a complete storage gene is produced every time step.}

\end{enumerate}

With these assumptions, we can achieve our target write speed of 100MB of data per second:

\begin{eqnarray*}
 \text{1 gene/\(200\ \mu \)s} & = & \text{5 genes/ms} \\ 
                                  & = & \text{50,000 symbols/ms} \\
                                  & = & \text{100,000 bytes/ms} \\
                                  & = & \text{100 megabytes/s}
\end{eqnarray*}

    
\section{DNA Assembly and Droplet Routing}
\label{sec:DNA Assembly}

\subsection{Automated DNA Assembly}
There are two requirements for our DNA synthesis system:

\begin{enumerate}
	\item To represent arbitrary data, it must assemble DNA oligos in any given order. 
	\item To assemble DNA oligos efficiently, several strands must be assembled simultaneously in one Gibson  process, without risking misalignment.
\end{enumerate}

These two requirements are contradictory. If oligos can come in \emph{any} order, one cannot join more than two simultaneously; they could join in the wrong order. To ensure desired ordering, it is necessary for segments to be uniquely matched with one another.

\begin{figure*}[h]
	\centering
	\includegraphics[scale=0.64]{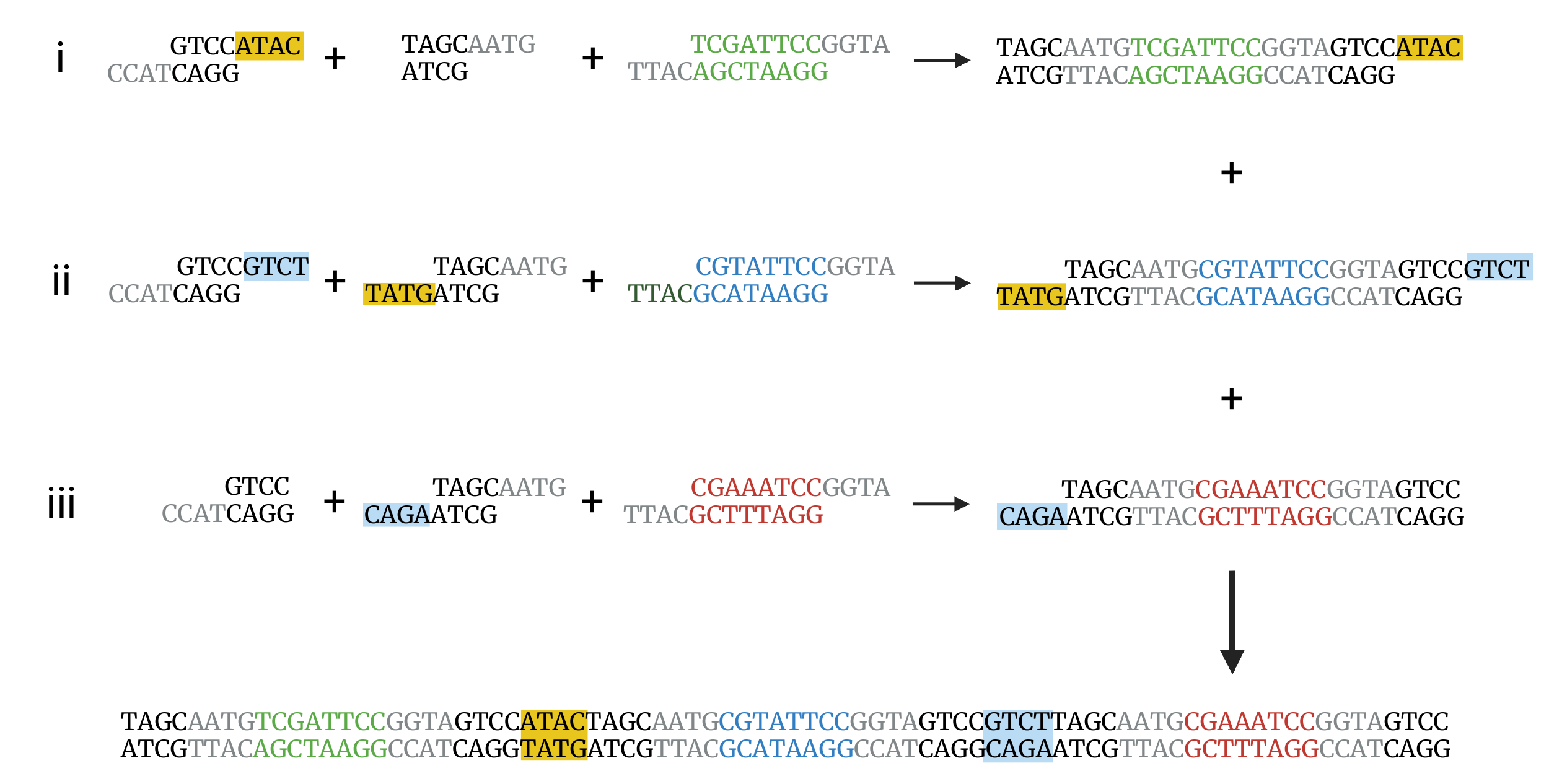}
	\caption{A graphical representation of the symbol-linker assembly process. A set (i-iii) of three symbols are joined to two linkers each in a first reaction.  The resulting one-symbol assemblies are assembled in a second reaction into a three-symbol assembly via the linkers. Attribution: this figure was created with the software \emph{Biorender}.}
	\label{fig:Symbol_Linker1}
\end{figure*}
\begin{figure}[h]
	\centering
	\includegraphics[scale=0.35]{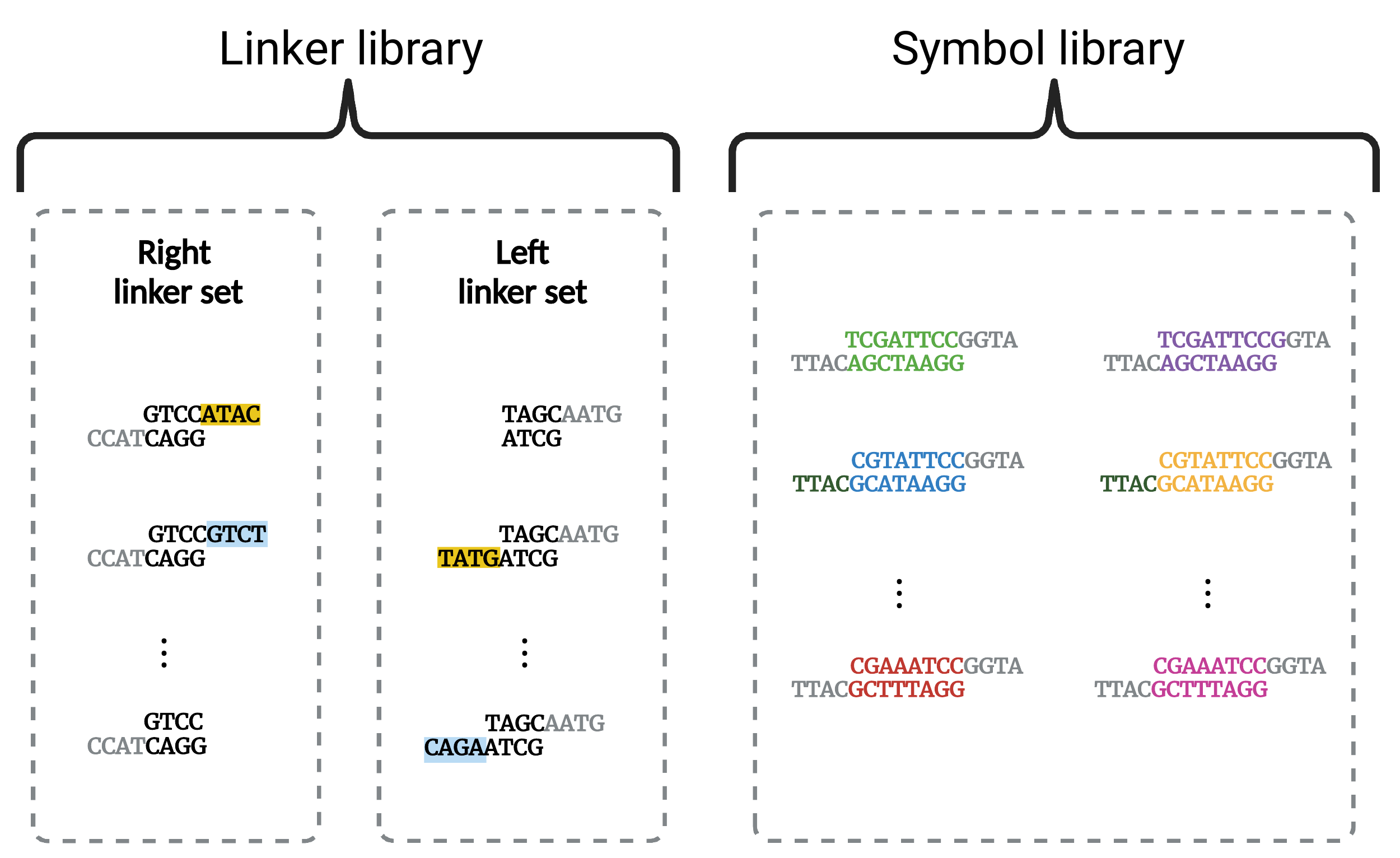}
   \caption{An example of a linker library and a symbol library.  The linker library contains two sets of linkers: a \emph{Right Linker Set} that attaches to the right end of a symbol, and a \emph{Left Linker Set} that attaches to the left end of a symbol. Universal overhangs (in gray) are used to attach any linker to any symbol. The highlighted regions of the linker sets are complementary to each other so that they can link together specifically during a Gibson reaction. Attribution: this figure was created with the software package \emph{Biorender}.}
	\label{fig:Libraries}
\end{figure}

We resolve this contraction with a dual library of oligos we call ``symbols'' and ``linkers'', illustrated in Fig. \ref{fig:Libraries}. The symbols allow us to represent arbitrary data when assembling them together. The linkers allow parallelization in the assembly process, ensuring correct ordering. 

Data genes comprise long chains of alternating symbols and linkers, with relevant information contained in the symbols only. All symbols have unique interior segments composed of 8 base pairs, allowing each to encode 16 bits. By using multi-bit symbols instead of assembling one base pair at a time, we exchange much of the fabrication time for overhead in maintaining the symbol library.

All symbols share the same beginning (\textit{left-side}) and end (\textit{right-side}) sequences. The left and right ends are not complementary, disallowing direct Gibson assembly of two symbols. Complementary ends are shared by all linkers, but each linker only has one end matched with those of the symbols. The other end binds with its unique, complementary linker. Thus, any desired chain of symbols can be assembled by first using Gibson assembly to separately attach each symbol to the appropriate linkers and then bringing all attached symbol-linker pairs together in another Gibson assembly process. 

The linkers will naturally order themselves according to their unique matches and the symbols will automatically fall into the appropriate order. This process is demonstrated in Fig. \ref{fig:Symbol_Linker1}. Following assembly, the new string of symbols undergoes purification and polymerase chain reaction (\textit{PCR})~\cite{newton1997pcr}. This product is a storage gene that holds encoded information in its symbols. 

Any gene requiring more symbols than what can be reliably handled by a single assembly process can be constructed by repeated assembly processes. Now, linkers on the ends of longer segments -- each consisting of multiple symbols -- specify the order in which these should be assembled. 

Assembling large data sets, consisting of millions, billions, or trillions of symbols, will require vast numbers of individual Gibson assembly operations. This presents a non-trivial problem in the form of managing droplet traffic routes and congestion. Given an arbitrary list of symbols to be encoded in a gene, droplets must be created, destinations chosen, and routes calculated. This multistep process necessitates an automated system capable not only of routing traffic but also deciding what Gibson assembly operations must be performed and when to build the desired gene.

\subsection{Routing Algorithm for Droplet Pathing} \label{sec:A_star}
Our system routes droplet traffic to desired Gibson, PCR, and purification sites using prioritized 3-D A* \cite{wang20193}  on the DMFB. First, generic A*\cite{bell2009hyperstar} will be explained, and then the prioritized 3-D A* algorithm that we use will be discussed in detail.

\begin{figure}[h]
	\centering
	\includegraphics[scale=0.61]{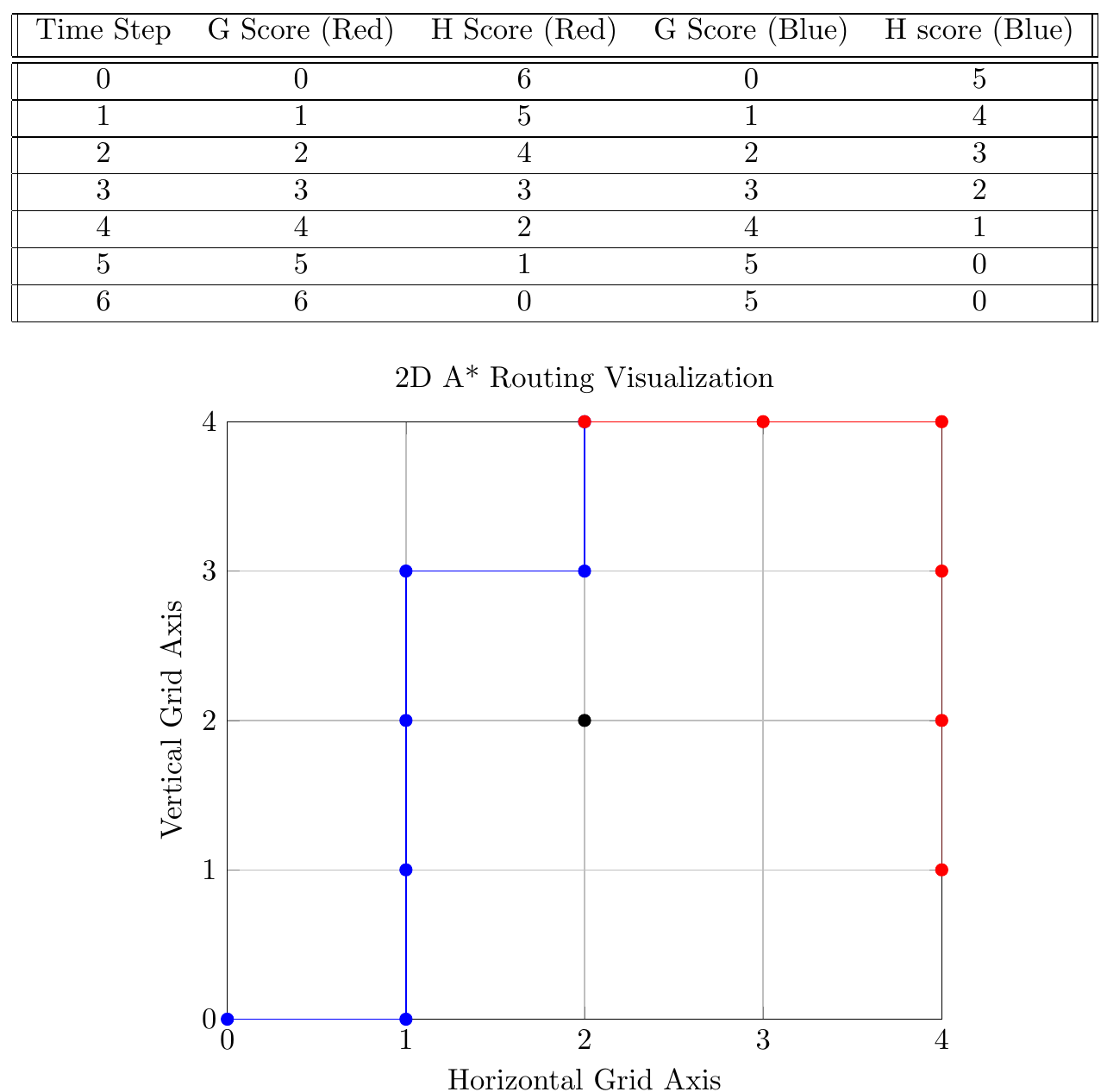}
	\caption{The top table shows the $f$ and $g$ scores for each path at every time step. The bottom figure illustrates 2-D A* with a simple example of 2 droplets shown in a 2-D plane. These 2 droplets wish to mix at the goal coordinate (2,4). The droplet corresponding to the blue route starts at (0,0), and the droplet corresponding to the red route starts at (4,1). Both droplets must consider an obstacle located at (2,2) while computing routes. The blue route will be calculated first due to the fact that it has the largest distance to the goal coordinate. The red route will then be calculated after. The table contains both the $g$ and $h$ scores of each path at each time step for the paths taken.}
	\label{fig:2dAStar}
\end{figure}

The goal of A* is to find the lowest cost path from point A to point B on a given graph. A graph is a generic collection of nodes connected via edges, and cost refers to the length of the path. The costs of paths are calculated using two scores, referred to as $g$ and $h$ scores. The $g$ score is the cost to get to the current node (the path already traversed), and the $h$ score is the distance from the current node to the end node (the path to traverse). In general, this $h$ score is determined via a specific user-chosen heuristic; our implementation of A* uses Manhattan distance. The $h$ and $g$ scores are added to become the $f$ score, or the total score for the path. Fig. \ref{fig:2dAStar} shows an example of A* on two droplets moving in 2-D space. Ideally, the $f$ score should be as small as possible \cite{zhang2020novel}.

Starting from point A, we look at each edge extending out from A and calculate the $f$ score for the surrounding nodes. The nodes and $f$ score are placed into an open set, usually represented as a data structure in memory. From there, paths are extended by looking at each node in the open set, starting with the lowest $f$ score. The $f$ scores for the nodes connected to the current node are then updated and placed back into the open set. This continues until B is reached, or it is concluded that no path to B is available.

To contextualize this algorithm, the nodes in the graph represent grid spaces on the DMFB, and edges indicate which grid spaces are next to each other. In the explanation below, point A represents a droplet's starting position while point B is its local destination which can be one of many things. It can be an intermediate location where the symbol and linker droplets mix into a larger droplet. It can also be a location where Gibson mixing occurs between the larger droplet and a Gibson mix reagent droplet. For both situations, this target location is the site of droplet mixing in some form. We classify the collection of these droplets as a \emph{merge} group.

\begin{figure}[h]
	\centering
	\includegraphics[scale=0.55]{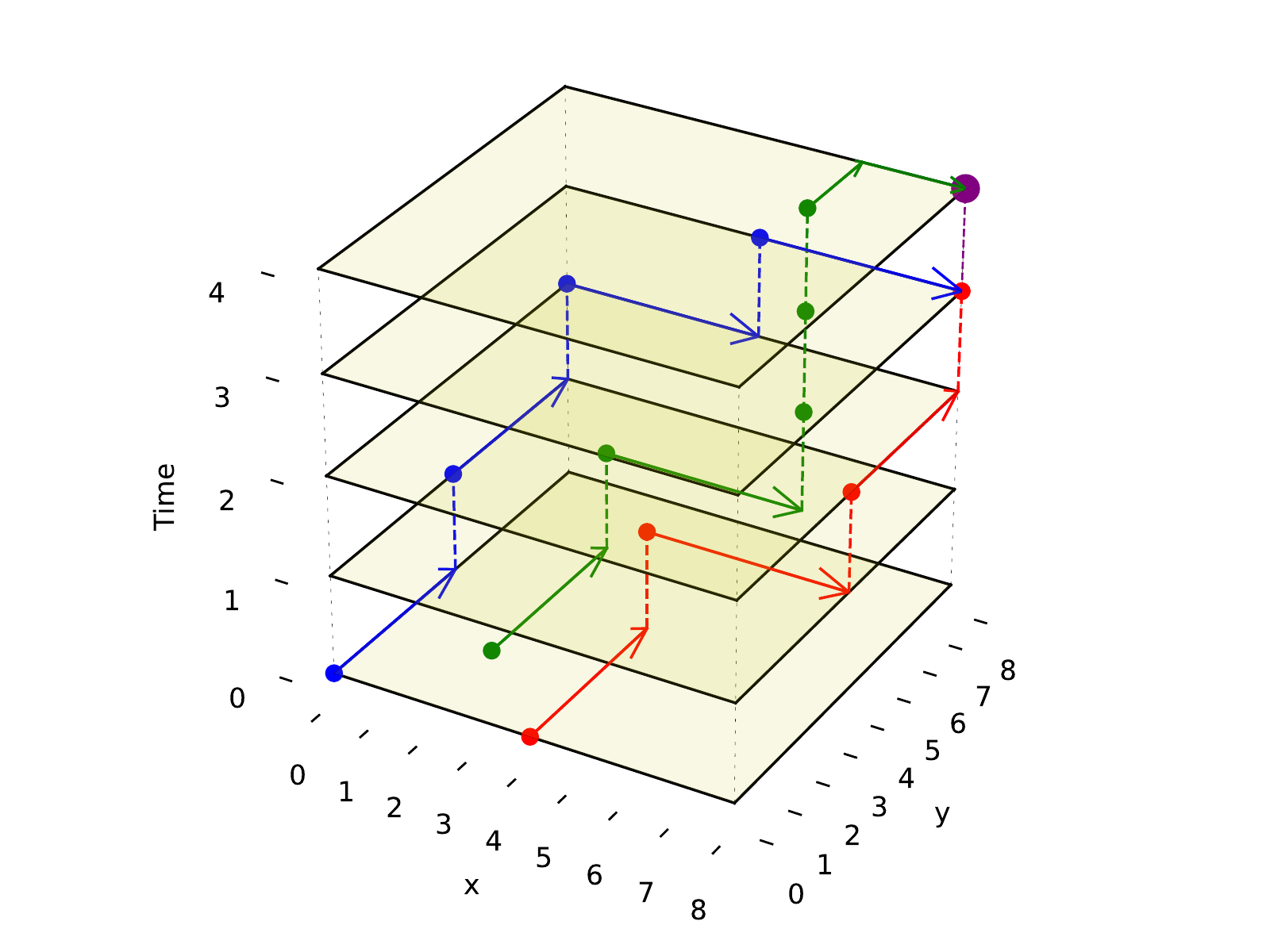}
	\caption{This figure illustrates 3-D routes with a simple example for 3 droplets. These 3 droplets wish to mix at the goal coordinate (8,8). The droplet corresponding to the blue route starts at (0,0,0), the droplet corresponding to the red route starts at (4,0,0) and the green droplet starts at (2,2,0). The movement of the droplet is shown by increasing the time value (z-axis value) by 1. The A* algorithm is projected to the 3-D space, and the blue route is planned first because it has the farthest Manhattan distance to travel. The routes are designed to avoid unwanted collisions with each other until they reach the desired location as no paths intersect. The merged droplet, which is bigger, is shown in purple at (8,8,4).}
	\label{fig:paths}
\end{figure}

We adapt the generic A* algorithm in our application to the prioritized 3-D A* form. It is prioritized because the algorithm tackles routing the droplet with the furthest Manhattan distance to travel first. It operates on 3 dimensions, with two dimensions representing the DMFB 2-D grid layout and the third representing time. All droplets are routed sequentially using the 3-D A* priority scheme. All droplets must move one grid space at a time simultaneously as all routes are planned beforehand. The algorithm is called $N_D$ times, where $N_D$ is the number of droplets on the grid. 3-D A* would be called once per newly pulled droplet. Once each droplet has its route, they will move together one time step at a time. The A* algorithm itself is called upon every merge operation and route completion as the resultant droplet now needs to be assigned a new route. Visualization of such routing movement is shown in Fig. ~\ref{fig:paths}

To prevent the unwanted merging of droplets, the notion of a droplet shadow and occlusion zone are introduced, as illustrated in Fig. \ref{fig:Shape1}. These are projected into a 3-D space such that for $N_D$ droplets, there are $3 N_D$ occlusion zones where each occlusion zone is present for the previous, current, and future time steps of the droplet. Droplets that are not in the merge group of the current droplet see the occlusion zones as obstacles they must route their paths around. Since routing is completed by taking all 3 dimensions into consideration and before any droplet movement occurs, the obstacles are static meaning they do not appear at random to block a droplet's route. This method of routing is advantageous because it prevents unwanted collisions from occurring while having each droplet take the optimal path given the constraints imposed by previously routed droplets.


\section{Architecture}
\label{sec:Architecture}

\begin{figure}[h]
	\centering
	\includegraphics[scale=0.5]{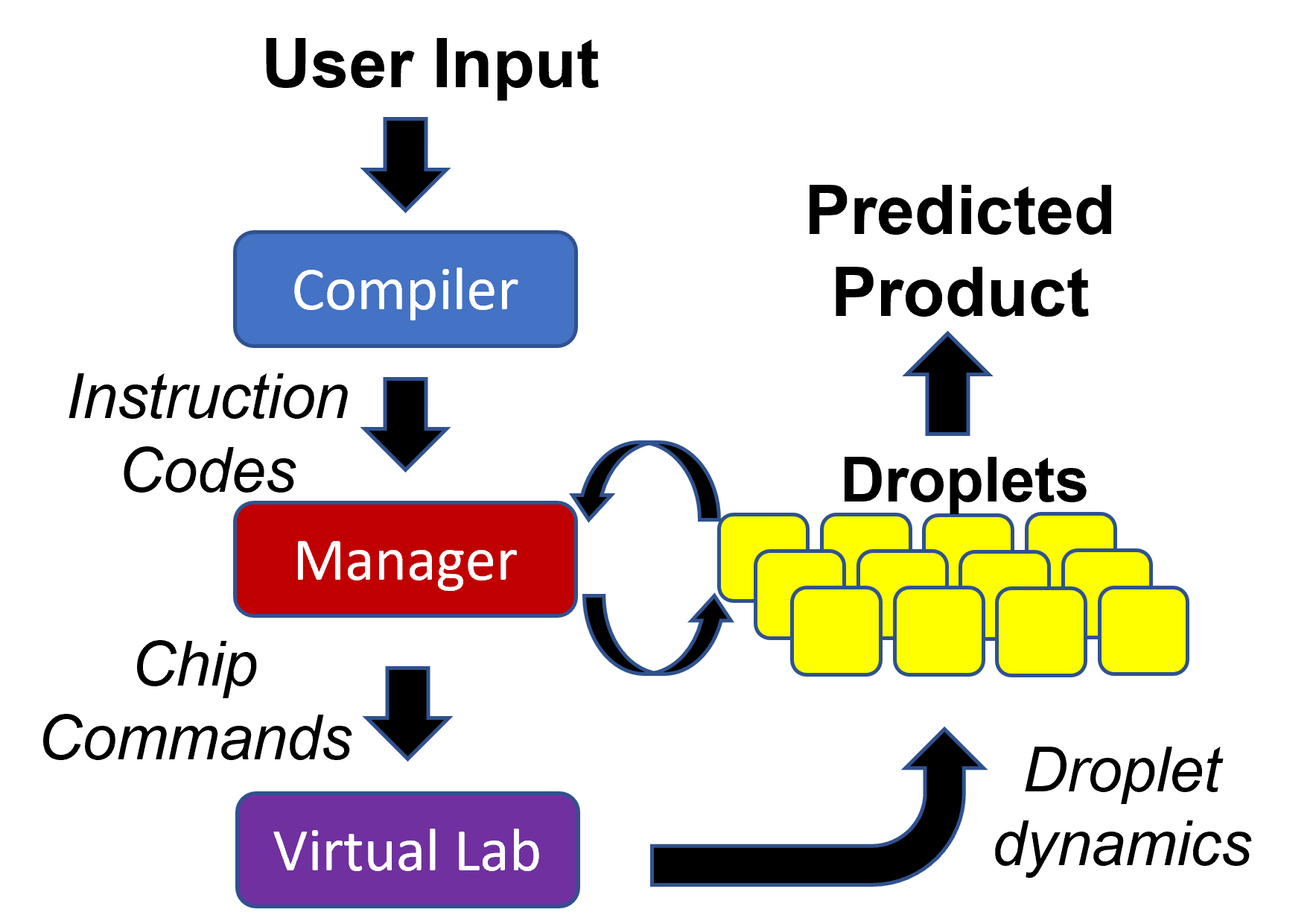}
	\caption{Block diagram of the software modules that plan and execute the automated gene assembly. The Compiler reads the desired input and consults a preprogrammed chemical protocol to generate the necessary chemical operations and properly order them. The Manager reads the resulting instruction codes and coordinates the creation, destination selection, and routing of droplets. The Virtual Lab  provides feedback on droplet movements to the manager.}
	\label{fig:Hierarchy1}
	\vspace{-0.25cm}
\end{figure}

We discuss the architecture of the software that controls the DMFB device. We use a modular hierarchy to solve the droplet traffic management problem, shown in Fig. \ref{fig:Hierarchy1}. At the top of the hierarchy, we have the Compiler, which is responsible for reading a user's desired data, in the form of a DNA storage gene $P$, and breaking down the basic chemical steps that must be performed to create it. These basic chemical steps are passed down to the Manager, which is responsible for assigning the droplets' destinations and issuing commands. At the bottom of the hierarchy is the Virtual Lab, which emulates a real DMFB. The lab houses a group of grid spaces and chemical droplet objects, which represent their physical equivalents and behave in similar ways within their virtual space.


Our Virtual Lab takes into account the time required for all droplet operations, including storing and retrieving droplets from reservoirs located at the perimeter of the grid. In our simulations, we assume that all the chemical reactions for DNA assembly complete in a single time step. While this assumption may or may not be realistic, chemical reactions generally complete quickly with femtoliter volumes. It is important to note that this ``real'' time is distinct from the ``runtime'' required for the 3-D prioritized A* algorithm. Of course, in the eventual production-level device, all routing must be done in real-time, so our routing algorithm must complete faster than the time taken by physical droplet routing.

\subsection{Assembly Protocol}

The system requires an assembly protocol to follow to automate the construction of user-determined strings of DNA. This protocol  specifies the sequence of chemical reactions that attaches multiple independent pieces of double-stranded DNA to one another. An instruction produced by the protocol takes the form of a list in the following format: \vspace{0.1cm}\\
\vspace{0.1cm}\noindent\mbox{}\hspace{2em} [$<$InstructionType$>$, $<$droplet1$>$, $<$droplet2$>$, \ldots] \\
where $<$InstructionType$>$ is a string. Each droplet is represented by a list of strings in the following format: \vspace{0.1cm}\\
\vspace{0.1cm}\noindent\mbox{}\hspace{2em} [$<$reagent1$>$, $<$reagent2$>$, \ldots, $<$reagentN$>$]  \\
containing $N$ reagents, although it is often the case that N = 1.

When reading the instructions, the manager will execute a case structure based on $<$InstructionType$>$ using component droplets matching the descriptions given by $<$droplet1$>$, $<$droplet2$>$, etc. An example instruction might be [`GibsonMove', [\_S0\_], [L1], [`Gibson-mix']], indicating that the manager should identify three droplets containing the symbol 0, the linker 1, and some Gibson mixing chemicals and bring them together on a suitable Gibson site.

An example of the instruction codes for a single assembly step using the Gibson symbol-linker protocol is given below. This list assembles the data string S1-S0-S2, corresponding to symbols numbered 1, 0, and 2, respectively, from the total list of symbols available.

\medskip
\begin{algorithmic}
	\STATE 1. [`Gibson-Move', [`L0'], [`\_S1\_'], [`Gibson-mix']]
	\STATE 2. [`Gibson', [`L0', `\_S1\_', `Gibson-mix']]
	\STATE 3. [`Gibson-Move', [`L1'], [`L2'], [`\_S0\_'], [`Gibson-mix']]
	\STATE 4. [`Gibson', [`L1', `L2', `\_S0\_', `Gibson-mix']]
	\STATE 5. [`Gibson-Move', [`L3'], [`\_S2\_'], [`Gibson-mix']]
	\STATE 6. [`Gibson', [`L3', `\_S2\_', `Gibson-mix']]
	\STATE 7.\hspace{1.1mm}[`Gibson-Move', [`L1\_S0\_L2'], [`L3\_S2\_'], [`\_S1\_L0'], [`Gibson-mix']]
	\STATE 8. [`Gibson', [`L1\_S0\_L2', `L3\_S2\_', `\_S1\_L0', `Gibson-mix']]
	\STATE 9. [`Purify-Move', [`\_S1\_L0L1\_S0\_L2L3\_S2\_'], [`Purify-mix']]
	\STATE 10. [`Purify', [`\_S1\_L0L1\_S0\_L2L3\_S2\_', `Purify-mix']]
	\STATE 11. [`PCR-Move', [`\_S1\_L0L1\_S0\_L2L3\_S2\_'], [`PCR-mix']]
	\STATE 12. [`PCR', [`\_S1\_L0L1\_S0\_L2L3\_S2\_', `PCR-mix']]
\end{algorithmic}
\medskip

In the list above, the first two steps create the symbol-linker droplet [`\_SI\_L0'] through Gibson moving and mixing steps. The first instruction has the instruction type 'Gibson-Move' with three droplets containing the linker 0, the symbol 1, and some Gibson Mix. It will move the droplets to an available Gibson site. The second step initiates Gibson mixing and assembly on the three droplets to form the chain [`\_SI\_L0']. Likewise, steps 3 to 6 produce the droplets [`L1\_S0\_L2'] and [`L3\_S2\_']. Steps 7 and 8 take these three larger droplets and perform Gibson assembly on them at a suitable Gibson site. This creates the droplet [`\_S1\_L0L1\_S0\_L2L3\_S2\_']. The final four steps take the final droplet to purify (clean) and to PCR (amplify) sites to create the final data string S1-S0-S2 held together with linkers.

\subsection{Compiler}


The Compiler is analogous to a software compiler which translates user input into a set of primitive instructions. The input consists of a list of characters representing the aforementioned ``symbols'' such as `S1-S0-S2'. The job of the compiler is to determine how to build the given DNA strand by repeated and recursive applications of its assembly protocol. We designate the desired final product $P$, with a length of $L_P$ symbols.

The compiler must first determine how to construct $P$ using assembly operations that can combine at most $N_A$ segments simultaneously, with the limit $N_A$ being set by the assembly protocol. In this case, $N_A$ is the reliability margin of the symbol-linker Gibson assembly. We employ an $N_A$-ary data tree to store the construction blueprint. The root node stores $P$. The compiler symbolically breaks $P$ up into $N_A$ separate segments and stores each segment in a child node below the root. These segments are broken up in the same way, with new nodes storing the new, smaller segments. The tree is built from the bottom up until the final nodes contain segments of one symbol in length. This abstract string-building is mimicked by the DNA strand assembly. 

Algorithm \#\ref{proc:data-partition} and Algorithm \#\ref{proc:build-assembly-tree} explain the algorithms for building the assembly tree step-by-step. (We note that not all details are included in the pseudocode given here.)

\begin{algorithm}
	\caption{Data Partitioning (inputs: data list, $N_A$)}
	\label{proc:data-partition}
	\begin{algorithmic}
		\STATE 1. If data length ($L_P$) does not exceed $N_A$, break data into singlets and return.
		\STATE 2. Otherwise, break data into $L_P/N_A$ $N_A$-tuplets.
		\STATE 3. Add a final, shortened tuplet for any data remainder.
		\STATE 4. Return list of tuplets.
	\end{algorithmic}
\end{algorithm}

\begin{algorithm}
	\caption{Build Assembly Tree (inputs: gene list, $N_A$)}
	\label{proc:build-assembly-tree}
	\begin{algorithmic}
		\STATE 1. Create a list of nodes, one for each symbol in the gene.
		\STATE 2. While node list length exceeds $N_A$, repeat 3--8.
		\STATE 3. Partition the node list using Algorithm \#\ref{proc:data-partition}.
		\STATE 4. Empty the node list.
		\STATE 5. For each sublist in the node partition, repeat 6--8.
		\STATE 6. Create a parent node above all nodes in the sublist.
		\STATE 7. Create instruction list for parent node using Process \#\ref{proc:generate-instructions}.
		\STATE 8. Append parent node to nodelist.
		\STATE 9. Once the node list is less than $N_A$ nodes long, create a root node above all nodes remaining in the list.
		\STATE 10. Create an instruction list for the root node using Algorithm \#\ref{proc:generate-instructions}, then return.
	\end{algorithmic}
\end{algorithm}

\begin{figure}[h]
	\centering
	\includegraphics[scale=0.5]{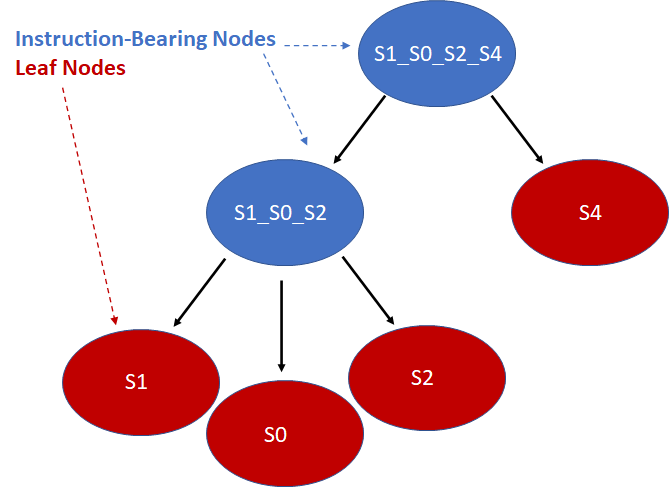}
	\caption{Simplified diagram of a small assembly tree data structure. In red, the leaf nodes each contain a single symbol. The root node represents the final desired symbol sequence or gene. Non-leaf nodes contain assembly instructions readable by the Manager. We note that the linkers between the symbols have been excluded for readability. See Fig. \ref{fig:Node1} for more details on individual nodes.}
	\label{fig:Tree1}
	\vspace{-0.2cm}
\end{figure}

With the ordering of the tree determined, the compiler will populate the instructions for each node by consulting the chemical protocol and giving the strands of a node's children as its inputs. This is outlined in Algorithm \#\ref{proc:generate-instructions}. This implies nodes without children (leaf nodes) have no instructions. The resulting `assembly tree' provides a blueprint for constructing the final product $P$. The leaf nodes, each holding one symbol, can be read from left to right to give the individual symbols of $P$. The first layer of non-leaf contains the $N_A$-length products of the first set of assembly operations as well as the instructions needed to carry them out. The next layer further groups those segments in length $N_A^2$, and so on. The root node contains instructions for the final assembly of $P$, grouping the remaining segments together. Fig. \ref{fig:Tree1} illustrates the assembly tree data structure used to assemble `S1\_S0\_S2\_S4' with linkers omitted. The instruction codes for assembling `S1-S0-S2', described above in the Assembly Protocol subsection, are carried by the left-hand instruction node in the graphic.

\begin{algorithm}[h]
	\caption{Generate Instructions (inputs: node, linker-list)}
	\label{proc:generate-instructions}
	\begin{algorithmic}
		\STATE 1. Create a list of subgenes in node's children.
		\STATE 2. For each subgene, repeat 3--6.
		\STATE 3. Select 3' and 5' linkers, if any.
		\STATE 4. Create an empty instruction list.
		\STATE 5. Add a `Gibson-Move' command to the instruction list with non-mixed subgene and linkers as variable reagents.
		\STATE 6. Add `Gibson' command to the instruction list with mixed subgene and linkers as variable reagent.
		\STATE 7. Add `Gibson-Move' command to the instruction list with non-mixed subgene-linker sets as variable reagents.
		\STATE 8. Add `Gibson' command to the instruction list with mixed subgene-linker set as variable reagent, then return.
	\end{algorithmic}
\end{algorithm}

Besides encoding the organization of substrings and the individual instructions necessary to chemically assemble $P$, the nodes also interface with the manager in real-time to track the disposition of droplets created for each node. This will be discussed in more detail below.

\subsection{Virtual Lab}



The Virtual Lab simulates each droplet and each DMFB grid space as independent objects in a continuous loop representing the passage of real-time. At each time step the lab checks for any update commands, activating or deactivating the corresponding grid spaces, if any. Then each droplet checks its surroundings for active grid spaces and updates its location according to the droplet movement model. For convenience, all droplet objects maintain references to each grid space they currently contact, and each grid space object similarly holds references to any and all droplets touching it. This location update step is where the lab detects errors.

\begin{figure}
	\centering
	\includegraphics[scale=0.5]{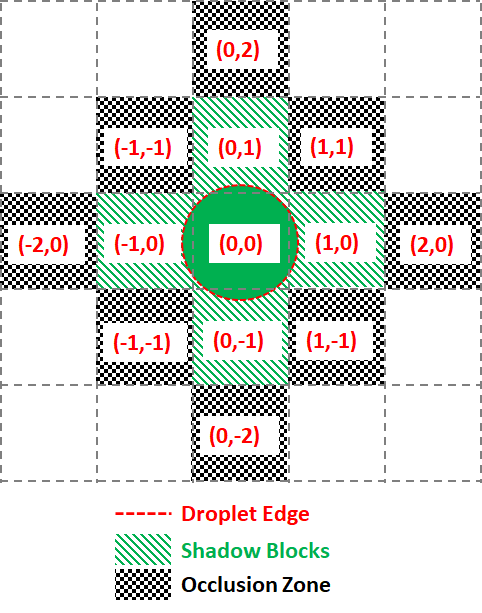}
	\caption{Diagram of the digitization of droplet shadow and occlusion zones. A droplet centered on a grid space with relative coordinates (0,0) has a sufficient radius to touch the four nearest neighboring grid spaces, making five shadow blocks. All grid spaces adjacent to the shadow blocks are designated as occlusion zones.}
	\label{fig:Shape1}
\end{figure}

The droplets are tracked in terms of their current `shadow', which is a digitization of the droplet's shape. Assuming a droplet is centered in the middle of a grid space, the shadow is a list of grid space coordinates, relative to the center space, which also touch the droplet. This is used by the manager to determine which electrodes to use for moving the droplet. This also allows easy calculation of an occlusion zone, the layer of grid spaces around a droplet that is as close as possible without touching. This layer is used as a barrier, off-limits to all other droplets that are not intended to mix. For instance, a small droplet that only touches the grid spaces nearest to its center in the four cardinal directions would have a shadow S = [[0,0], [1,0], [0,1], [-1,0], [0,-1]] and an occlusion zone O = [[1,1], [1,-1], [2,0], [0,2], [-1,1], [-2,0], [-1,-1], [0, -2]] as shown in Fig. \ref{fig:Shape1}. As droplets merge and grow, their shadows and occlusion zones increase commensurately. During routing, a droplet's shadow and occlusion zone are projected both forward and backward in time by one step to ensure no undesired mixing can happen.



\subsection{Manager}
\label{Manager}
With a simulated lab to house the droplets and a compiler to provide the basic mix and merge instructions, there is one final task. The instructions must be translated into actual commands for the lab to execute. This entails selecting reservoirs from which to pull droplets of appropriate types, choosing destinations for them, and determining routes that will see them to their destinations without any unwanted misadventures along the way.  This is the manager's job. The manager interfaces with the assembly tree and the lab. It runs in a loop matching the lab's time steps, providing new commands at each step while also tracking long-term progress toward each assembly node's instructions.

\begin{figure}[h]
	\centering
	\includegraphics[scale=0.42]{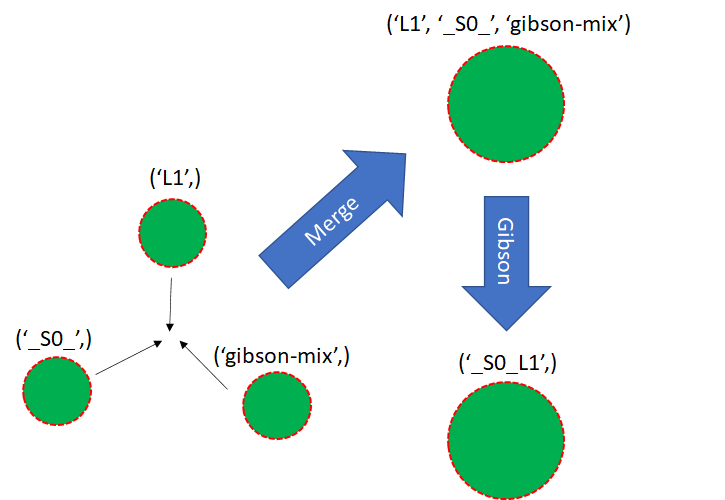}
	\caption{Three single-species droplets merge, becoming a larger droplet with mixed species. The droplet undergoes Gibson assembly, becoming a different species.} 
	\label{fig:Merging1}
\end{figure}

The manager must be able to track the droplets in the lab, knowing their locations and contents at each step. It is useful to reference droplets by their contents rather than location since this allows easy matching of droplets to instructions that call for specific reagents. 
Fig. \ref{fig:Merging1} visualizes the changes in droplet references after merging and performing Gibson assembly. 

\begin{figure}[h]
	\centering
	\includegraphics[scale=0.4]{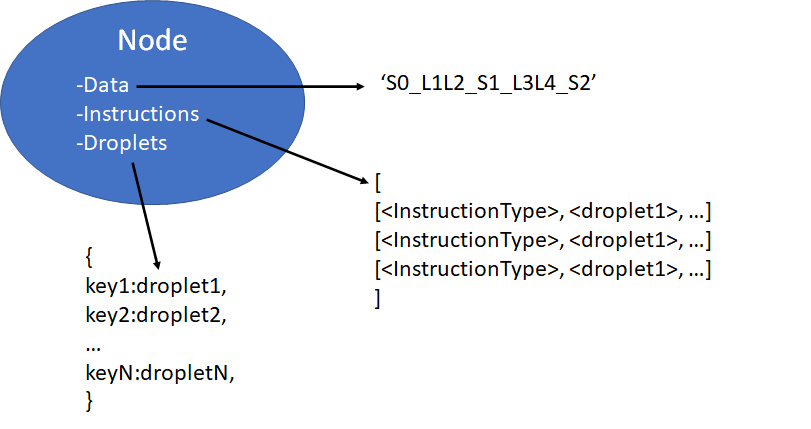}
	\caption{Detailed contents of an assembly tree node object. Three items are carried by each non-leaf node. The data string corresponds to the sequence of symbols and linkers produced by the node. Instructions, a list of the requisite fluidic and chemical operations. Droplets, a mutable hashmap of currently extant droplets being used for instructions by this node on the virtual lab.}
	\label{fig:Node1}
\end{figure}

We now have a complete description of the information contained in the Node objects.
As shown in Fig. \ref{fig:Node1}, they contain immutable $data$ and $instruction$ values consisting of the symbol-linker representation of the DNA strand they construct and the instructions used to build it. They also contain a mutable dictionary of droplet objects which changes throughout the manager-lab time loop.

The manager runs in a loop lock-stepped with the lab, maintaining a list of active nodes drawn from the assembly tree and stepping through their instructions in parallel. It also creates a list of lab commands which begins each iteration empty, fills up during the node advance and droplet routing steps, and is subsequently passed to the lab for execution. For each iteration, it runs four processes in order as shown in Algorithm\#\ref{proc:iteration}. Initially, the active nodes list consists of all the lowest level non-leaf nodes.

\begin{algorithm}[h]
	\caption{Complete one time step iteration (inputs: node)}
	\label{proc:iteration}
	\begin{algorithmic}
		\STATE 1. Check Node Progress.
		\STATE 2. Advance Node Instructions.
		\STATE 3. Plan Routes for Unrouted Droplets.
		\STATE 4. Execute One Round of Commands.
	\end{algorithmic}
\end{algorithm}

When checking node progress, the system evaluates the node's current instruction list and its droplet's location and sees if the node is ready to move to the next instruction. If the node is ready to advance its instructions, it will check the instruction codes and advance to the next instruction.

After finishing with the nodes, the manager checks for droplets that have been newly assigned to destinations and those that could not be routed during the last iteration. The manager attempts to plan a route for each of these.

The manager's routing algorithm uses the 3-D prioritized A* method discussed above in Section~\ref{sec:A_star}. Droplets are organized into `merge groups', which are collections of droplets that are seeking to combine into one conglomerate. All routes that are generated obey the following anti-collision constraint that applies to any two droplets $d_x$ and $d_y$ of different merge groups. Any grid space occluded by $d_x$ until time $t_a$ may not be overshadowed by $d_y$ during any time $t_b$ such that $t_a$ $\ge$ $t_b$. 

With priority given to merge groups containing droplets with the farthest Manhattan distance to travel, all selected droplets are routed one at a time using a 3-D A* graph traversal. Two of the dimensions represent the virtual--or physical--DMFB grid layout, while the third is time. 

For each droplet $d$ the router takes droplet shadow S$_d$ and occlusion zone O$_d$ into account at every step, projecting them into the 3-D space. These zones are off-limits for other droplets during their own routing phase. This includes droplets that will be routed during this or any future time steps. The space is initially free of occlusion zones when the highest-priority droplets are routed and becomes more populated as the other droplet routes are filled in. Of course, there may also be occlusion zones generated by the routing phase in the previous time step, which all droplets during this time step must avoid regardless of their priority level. There will be at most $3 N_D$ occlusion zones present on the grid at a given time, where $N_D$ is the total number of extant droplets. This is because each droplet generates occlusion zones for its most recent, current, and immediately subsequent steps to satisfy the anti-collision constraint. This routing method allows many droplets to move simultaneously while avoiding unwanted collisions. Furthermore, each individual droplet takes the optimal path within the constraints given to it by the droplets higher in the priority queue and the droplets routed during previous time steps.

We note that the modularity of the system makes it relatively easy to implement a different routing scheme.  Incorporating more advanced electrowetting technology, which would allow for more flexible movement, would only require the redesign of the routing subroutine itself, leaving the other parts of the software to function as normal. One possible redesign to routing is to make it contamination aware~\cite{huang2009contamination}. Another possibility includes incorporating the algorithm ``Moving Target D* Lite''~\cite{sun2010moving}, which has been shown to be effective in problems where obstacle occurrences appear dynamically over time. 

After planning routes for all droplets, the manager performs the movement of droplets according to the instructions and routes that have been set up. After these steps are complete, one loop iteration is concluded.



\section{Results and Discussion}
\label{sec:Results}

\subsection{Simulation Results}

Having presented the software architecture for our automated DNA assembly system, we now present simulation results. We note that our simulation does not incorporate physical latencies present with DMFBs such as the electrode switching rate~\cite {hua2010multiplexed} and the delay with mixing operations~\cite{loveless2020performance}. Accordingly, a comparison of runtimes with other DNA synthesis technology is speculative, at best.

\begin{figure}[h]
	\centering
	\includegraphics[scale=0.5]{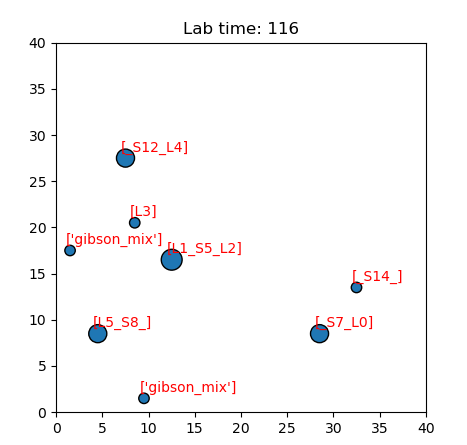}
	\caption{Image of the simulation's GUI displaying droplets being routed in a 40 $\times$ 40 size grid. The number of full command execution rounds is shown at the top as the lab time.}
	\label{fig:DMFSim1}
\end{figure}

In our simulation, we wish to evaluate the impact of computer hardware, virtual lab grid size, and target gene length on our system's performance. We draw conclusions about their impacts on the system's runtime, memory usage, and CPU usage. The simulated DMFB system for DNA assembly was written in Python. Fig. \ref{fig:DMFSim1} displays a snapshot of the simulation’s GUI displaying routed droplet movement in real-time. Additionally, benchmarking results were captured to better understand the distribution of function workload and the limitations of the current model.

\subsubsection{Hardware Results}

The program was executed on three different machines, each with different CPUs and different amounts of RAM. We used a constant grid size of 1,000 $\times$ 1,000 grid points (see Table \ref{t1} for machine specifications).\footnote{For our simulations,  we assume a grid size of one million points. However, as discussed in Section~\ref{sec:writespeed}, the eventual DMFB will have a much larger grid.} Table \ref{t1} displays the simulation’s runtimes in correspondence to hardware specifications. We note that RAM availability does not seem to be a bottleneck. The data gathered in Table \ref{t2} shows that peak RAM usage is approximately constant, about 1.3 Gigabytes (GiB), across all machines. 

\begin{table}[h]
	\centering
	\caption{RAM statistics vs. System Hardware. The RAM usage was approximately the same across the three machines when run on a grid size of 1,000 $\times$ 1,000 grid points. There are variations in total RAM, but all machines have enough RAM to support the memory required to run the simulation. Machine 3 had the fastest RAM data rate of 3,200 Mega Transfers per second (MT/s), and the other two machines had RAM data rates of 2,400 MT/s.}
	\begin{tabular}{|c|c|c|c|}
		\hline
		Machine & Peak RAM Used & Total RAM & RAM Speed \\
		\hline\hline
		1       & 1.4GiB        & 16GiB     & 2400MT/s  \\
		\hline
		2       & 1.3GiB        & 32GiB     & 2400MT/s  \\
		\hline
		3       & 1.2GiB        & 32GiB     & 3200MT/s  \\
		\hline
	\end{tabular}

	\label{t2}
\end{table}

The simulation consumes an entire CPU thread on each of the three machines. CPU usage is clearly the constraining factor of the simulation. We analyzed the CPU efficiency using a metric called ``thread rating,'' which is a metric that represents the performance of a single CPU core. As the thread rating increases, the runtime decreases in an exponential relationship. 

\begin{table}[h]
	\centering
	\caption{Simulation Runtime vs. System CPU Hardware. Thread rating is a metric obtained from PassMark Software's extensive database of CPU benchmarks and tests. The thread rating indicates the performance of a single logical CPU core. The units for the thread rating are $MOps/Sec$~\cite{performancetest}.}
	\begin{tabular}{|c|c|c|c|}
		\hline
		Machine & Thread Count & Thread Rating & Runtime (s) \\
		\hline\hline
		1       & 8            & 1903          & 7234.69     \\
		\hline
		2       & 8            & 2417          & 3191.83     \\
		\hline
		3       & 16           & 3143          & 1962.33     \\
		\hline
	\end{tabular}
	\label{t1}
\end{table}

The CPU usage results are summarized in Table \ref{t1}.  Since Machines 1 and 2 have 8-thread CPUs, they yielded around 12.5\% CPU usage. Meanwhile, Machine 3 has a 16-thread CPU yielding around 6.25\% CPU usage.

\begin{figure}[h]
	\centering
	\vspace{5pt}
	\begin{tikzpicture}[scale = 0.8]
		\begin{axis}[scaled y ticks=false,
				nodes near coords,
				point meta=explicit symbolic,
				title={},
				xlabel={Thread Rating (MOps/sec)},
				ylabel={Runtime (sec)}]
			\addplot table [meta index=0, x=tr, y=rt, col sep=comma] {thread-rating-v-runtime.csv};
		\end{axis}
	\end{tikzpicture}
	\caption{Simulation Runtime vs. Thread Rating. Data points are labeled by the corresponding machine number, see \ref{t1}. As the thread rating increases, the runtime decreases. Intuitively there is a diminishing return on decreasing runtime with an increase in thread rating. A non-asymptotic relationship between thread rating and runtime would imply computations could be done instantaneously with substantial thread ratings.}
	\label{fig:RunVsThread}
\end{figure}
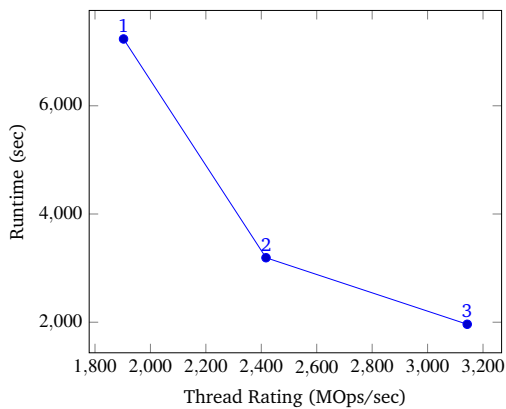

\subsubsection{Grid Size Results}

\begin{figure}[h]
	\centering
	\vspace{5pt}
	\begin{tikzpicture}[scale = 0.8]
		\begin{axis}[title={}, xlabel={Grid size (n x n)}, ylabel={Runtime (sec)}]
			\addplot table [x=g, y=t, col sep=comma] {gs-v-runtime.csv};
		\end{axis}
	\end{tikzpicture}
	\caption{Runtime vs. Grid size (n x n). The runtime of the simulation synthesizing a gene of length 5 was captured for grid axes lengths of 500 to 1500. There is a clear exponential increase in time with a linear increase in grid size.}
	\label{fig:RunVsGrid}
\end{figure}
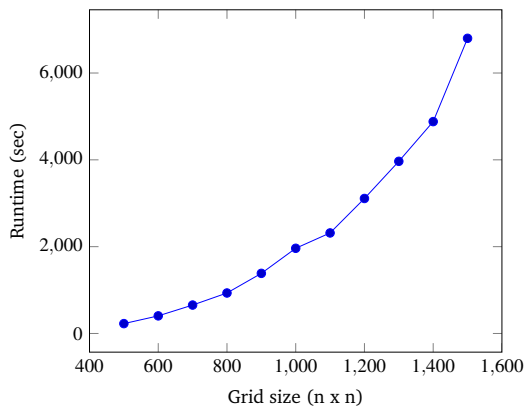

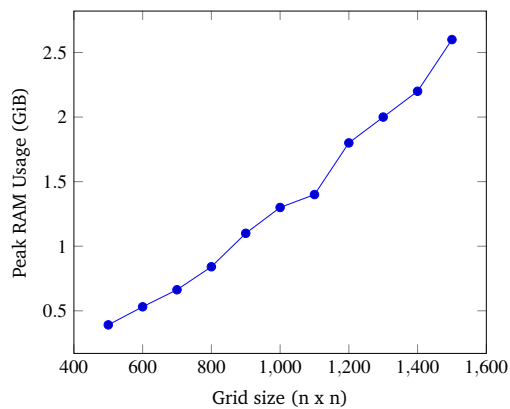
\begin{figure}[h]
	\centering
	\vspace{5pt}
	\begin{tikzpicture}[scale = 0.8]
		\begin{axis}[title={}, xlabel={Grid size (n x n)}, ylabel={Peak RAM Usage (GiB)}]
			\addplot table [x=g, y=m, col sep=comma] {gs-v-mem.csv};
		\end{axis}
	\end{tikzpicture}
	\caption{Peak RAM Usage vs. Grid size (n $\times$ n). The peak RAM usage (in GiB) of the simulation synthesizing a gene of length 5 was captured for grid axes lengths of 500 to 1500. There is a clear linear increase in memory consumption with a linear increase in grid size.}
	\label{fig:RAMVsGrid}
\end{figure}

\begin{figure}[h]
	\centering
	\vspace{5pt}
	\begin{tikzpicture}[scale = 0.8]
		\begin{axis}[scaled y ticks=false,
				yticklabel=\pgfkeys{/pgf/number format/.cd,fixed,precision=2,zerofill}\pgfmathprintnumber{\tick},
				ymin=6.1, ymax=6.3,
				title={},
				xlabel={Grid size (n x n)},
				ylabel={\% CPU Usage}]
			\addplot table [x=g, y=c, col sep=comma] {gs-v-cpu.csv};
		\end{axis}
	\end{tikzpicture}
	\caption{Average CPU Usage vs. Grid size. The Average CPU usage of the simulation synthesizing a gene of length 5 was captured for grid axes lengths of 500 to 1500. For all grid sizes, an average of 99\% of a single thread was consumed. Because the machine has 16 threads, the results showed slightly less than 6.25\% of CPU used across all grid sizes. There is no meaningful impact on CPU usage from changed grid sizes.}
	\label{fig:CPUVsGrid}
\end{figure}
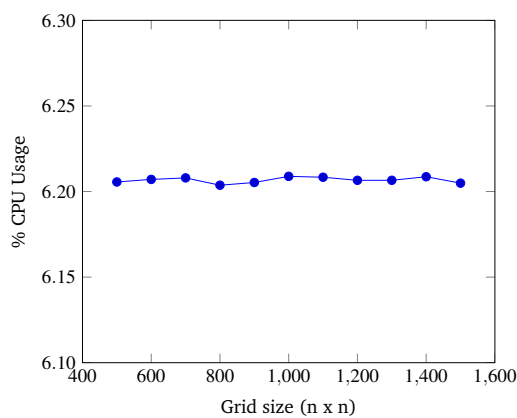

The program's performance was evaluated across a range of grid sizes on a single machine. Here, we used Machine 3. With respect to grid size, runtime responded exponentially (Fig. \ref{fig:RunVsGrid}), memory usage responded linearly (Fig. \ref{fig:RAMVsGrid}), and CPU usage responded constantly (Fig. \ref{fig:CPUVsGrid}). For very large grid sizes, RAM availability will become the limiting factor to affect runtime as a linear increase in grid size yields a linear increase in Peak RAM usage. Once Peak RAM usage nears the maximum available RAM, performance will deteriorate substantially. The CPU metrics do not have much impact on the runtime when sweeping across large grid sizes as they continue to consume the entirety of one thread.

\subsection{High Impact Functions}

\begin{table}[h]
	\centering
	\caption{Local runtime trials. The table illustrates the rationale behind the parameters used in each trial displayed in Fig. \ref{fig:LocalRuntimes}. The parameters for these trials were chosen such that the local runtimes of the simulation's functions could be examined as the problem size increases, but the congestion remains constant. In this situation, congestion is measured as the ratio of the total number of droplets pulled from reservoirs to the number of grid points. Droplet counts and grid sizes are discrete, thus the congestion is only approximately constant.}
	\begin{tabular}{|c|c|c|c|c|}
		\hline
		Trial & Grid Size & Gene Length & Droplets & Congestion \\
		\hline\hline
		1     & 50        & 2           & 9        & $0.00360$  \\
		\hline
		2     & 76        & 4           & 21       & $0.00364$  \\
		\hline
		3     & 96        & 6           & 33       & $0.00358$  \\
		\hline
	\end{tabular}

	\label{t4}
\end{table}

\begin{figure}
	\centering
	\vspace{5pt}
	\begin{tikzpicture}
		\begin{axis}[
				ybar stacked,
				ymin=0,
				enlarge x limits={abs=40pt},
				bar width=30pt,
				nodes near coords,
				every node near coord/.append style={xshift=-15pt,anchor=east,font=\footnotesize},
				legend style={at={(0.5,-0.20)},
						anchor=north,legend columns=-1},
				ylabel={Runtime distribution by function (\%)},
				xlabel={Grid Size (n x n)},
				symbolic x coords={50, 76, 96},
				xtick=data,
				legend style={at={(0.5,-0.2)}, anchor=north,legend columns=-1}
			]
			\addplot+[ybar] plot coordinates {(50, 11.0) (76, 79.5) (96, 82.1)};
			\addplot+[ybar] plot coordinates {(50, 1.3) (76, 6.5) (96, 9.7)};
			\addplot+[ybar] plot coordinates {(50, 87.7) (76, 14.0) (96, 8.2)};

			\legend{Advance  , Route Droplets  , Other  }
		\end{axis}
	\end{tikzpicture}
	\caption{Runtime distribution by major function. This data was obtained by running the simulation at an approximately constant congestion ratio. Here the congestion ratio is the total number of droplets pulled from reservoirs divided by the number of grid points. The approximate constant congestion ratio used was $0.0036$ droplets/gridsize$^2$  which was chosen by convenience. The parameters for these trials are shown in \ref{t4}.}
	\label{fig:LocalRuntimes}
\end{figure}
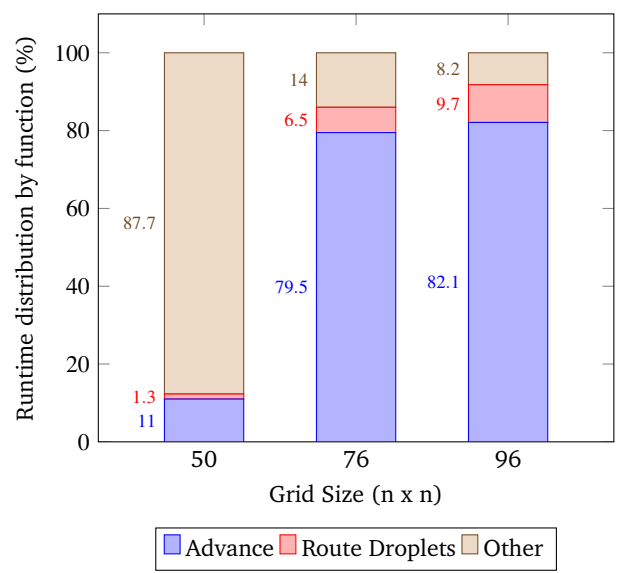

The open-source visualization software \emph{Gephi} was used to record the runtimes of the simulation's local functions over three trials. The trials were chosen such that as the problem size increased, the congestion remained constant. In the context of these trials only, congestion is computed as  the total number of droplets pulled from reservoirs divided by the number of total grid points. The exact simulation input parameters are displayed in Table \ref{t4}.

In these trials, the functions \verb+Advance+ and \verb+Route_Droplets+ were identified as methods of interest. The lab's function \verb+Advance+ computes an entire time step of the simulation. \verb+Advance+ iterates over the grid points and droplets multiple times in order to update their contents. The manager's function \verb+Route_Droplets+ computes the routes for droplets that have yet to be routed with the prioritized 3-D A* routing algorithm. The proportion of the total runtime spent computing each function is shown in Fig.~\ref{fig:LocalRuntimes}. As the problem size increases, \verb+Advance+'s proportion of runtime increases. The proportion of the time consumed by \verb+Route_Droplets+ also increases as the problem size increases, overtaking the combined runtime of all other subroutines. In the last trial, it is evident that these two functions will dominate the share of the simulation's runtime as larger, more realistic input parameters are chosen. It may be inferred that the computation of \verb+Advance+ on an exponentially increasing input tends to be slower than the computation of the prioritized 3-D A* routing algorithm on a linearly increasing number of pulled droplets.

Improving the runtime of \verb+Advance+ is worth attention in future work. This task is difficult as it requires an optimized approach to iterating over all grid points and droplets. Additionally, this is a function that runs on the simulated lab, and it is likely the speed of advancing droplets on a real DMFB will be different. The simulation may benefit in runtime by considering alternatives to the prioritized 3-D A* routing algorithm used in \verb+Route_Droplets+. Currently, the algorithm considers all possible routes for each droplet; however, there may be room for improvement by implementing methods that return an acceptable suboptimal route while only evaluating a fraction of the input space. Finding an acceptable suboptimal route for problems that can face a lot of congestion and time consumption has been studied by literature and shown to be effective in similar situations~\cite{8530894,katoch2021review}. These algorithms define an acceptable threshold for a path to be executed and then create the route once a path is found that meets the threshold limit.

\subsection{Limitation Testing}
This section explains the input spaces of interest where the program may face serious bottlenecks or failures at a high level. The inputs used to test the simulation specify a random gene consisting of 5 symbols at the expected grid size of 1000 $\times$ 1000 grid points. When the input grid size exceeded this grid size the program's runtime increased beyond practicality, taking over an hour to synthesize the gene routing about 30 droplets.

On the other hand, the program fails when grid sizes are small enough to generate considerable congestion. Within the context of the simulation, when the grid size becomes smaller than 40 $\times$ 40, given all other default parameters, there are issues with synthesis due to droplet congestion. A gene of length 5 pulls a total of 29 droplets as shown in Fig. \ref{fig:TotDropVsMaxDrop}, and all of these droplets need to be routed without causing collisions. In these situations, droplet paths may be blocked long enough to trigger a timeout where the synthesis of the gene is no longer pursued.

Aside from grid size, the number of reaction sites (Gibson, purification, PCR) can limit runtime and potentially cause timeout from congestion if very few of these sites exist on the chip. Moreover, there is a limited amount of chemistry sites the DMFB will allow due to its physical grid size. The number of reaction sites the DMFB contains depends on its physical grid size. Additionally, the number of reaction sites cannot exceed the number of possible reaction site locations.

\subsection{Congestion Testing}

\begin{figure}[h]
	\centering
	\includegraphics[scale=0.3]{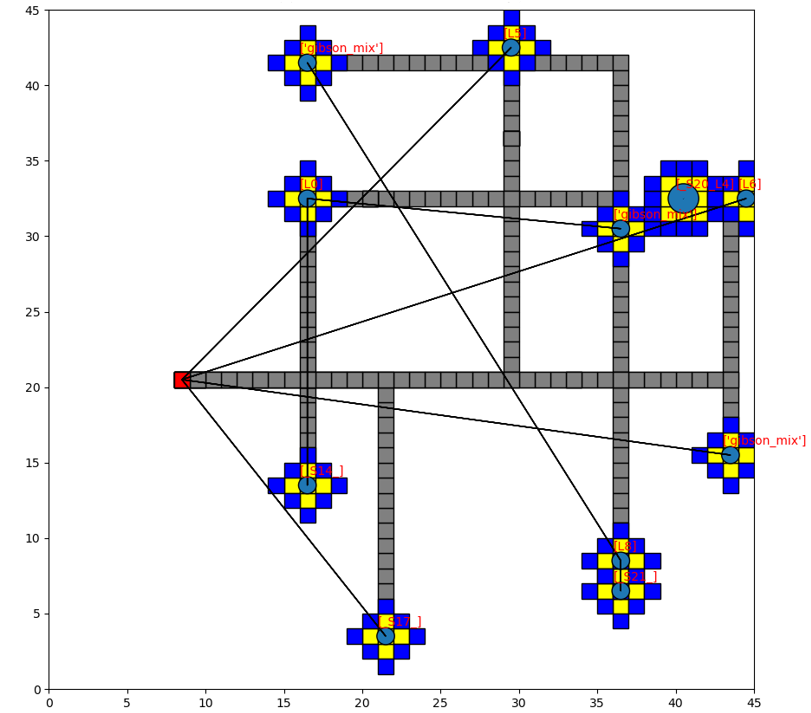}
	\caption{Image of the simulation's GUI displaying droplets being routed in a 45 $\times$ 45 size grid at a gene length of 10 with additional droplet information visualization. The number of full command execution rounds and congestion are shown at the top as the lab time. Congestion is computed as the number of routed (gray), shadow (yellow), and occluded (blue) grid points divided by the total number of grid points. The gray squares are included in congestion calculations as they are counted as used grid space for droplet routes. Congestion, in this sense, is a broad description of the simulated lab's congestion as a whole. Routed grid points are included in the calculation to indicate routing congestion for the prioritized 3-D A* algorithm.}
	\label{fig:DMFSim2}
\end{figure}

Fig. \ref{fig:DMFSim2} shows a visualization of multiple droplets being routed to one Gibson site. The yellow squares represent the shadow of a droplet and the blue squares are their occlusion zones. We display the paths of various droplets shown as gray squares to indicate these as ``no-go'' zones for other droplets. To other droplets, these gray squares are temporarily forbidden from crossing to avoid contamination of the droplet as it may pick up debris from a previously routed droplet, which is undesirable. Of course, this issue is hardware specific, but it is considered a constraint in our simulation and therefore contributes to congestion calculations.

The program can time-out under conditions of extreme congestion. This may occur for a number of input combinations, most notably when the grid size becomes too small. In these cases, there may be too few reaction sites, or too many droplets (as a consequence of synthesizing a long target gene). Any combination of these factors may lead to a situation where droplet paths are blocked and progress cannot be made. The simulation then times out and the synthesis of the target droplet is abandoned.

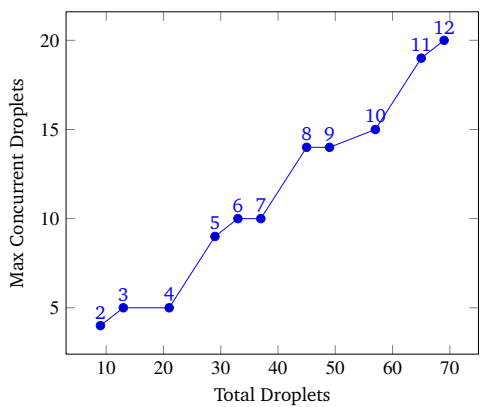
\begin{figure}[h]
	\centering
	\vspace{5pt}

	\begin{tikzpicture}[scale = 0.8]
		\begin{axis}[scaled y ticks=false,
				nodes near coords,
				point meta=explicit symbolic,
				title={},
				xlabel={Total Droplets},
				ylabel={Max Concurrent Droplets}]
			\addplot table [meta index=0, x=td, y=md, col sep=comma] {gl-vs-congestion.csv};
		\end{axis}
	\end{tikzpicture}
	\caption{Max Concurrent Droplets vs. Total Droplets. This data was obtained by running the simulation at a constant grid size of 45 $\times$ 45 over a range of gene lengths two to twelve. The number above each point represents the gene length. The vertical axis displays the maximum number of droplets recorded on the chip during the entirety of the simulation while the horizontal axis displays the total amount of droplets pulled from reservoirs to synthesize the gene.}
	\label{fig:TotDropVsMaxDrop}
\end{figure}

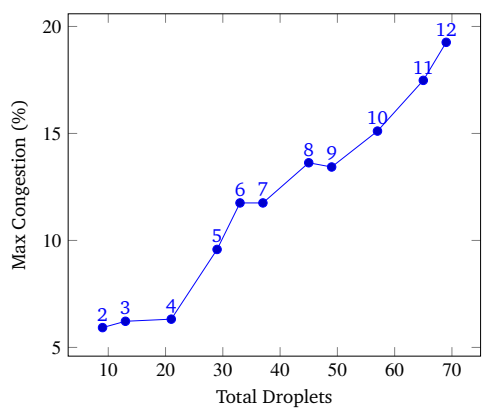
\begin{figure}[h]
	\centering
	\vspace{5pt}

	\begin{tikzpicture}[scale = 0.8]
		\begin{axis}[scaled y ticks=false,
				nodes near coords,
				point meta=explicit symbolic,
				title={},
				xlabel={Total Droplets},
				ylabel={Max Congestion (\%)}],
			\addplot table [meta index=0, x=td, y=mc, col sep=comma] {gl-vs-congestion.csv};
		\end{axis}
	\end{tikzpicture}
	\caption{Total Droplets vs. Max Congestion. This data was obtained by running the simulation at a constant grid size of 45 $\times$ 45 over a range of gene lengths two to twelve. The number above each point represents the gene length. The vertical axis displays the total congestion and the horizontal axis displays the total amount of droplets pulled from reservoirs to synthesize the gene. The relationship between the number of pulled droplets and the maximum congestion is approximately linear.}
	\label{fig:TotDropVsMaxCong}
\end{figure}

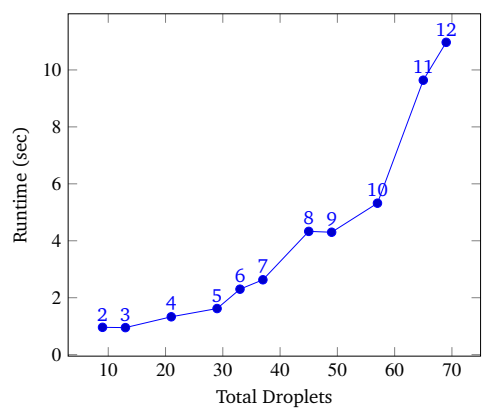
\begin{figure}
	\centering
	\vspace{5pt}
	\begin{tikzpicture}[scale = 0.8]
		\begin{axis}[
				scaled y ticks=false,
				nodes near coords,
				point meta=explicit symbolic,
				title={},
				xlabel={Total Droplets},
				ylabel={Runtime (sec)}
			]
			\addplot table [meta index=0, x=td, y=rt, col sep=comma] {gl-vs-runtime.csv};
		\end{axis}
	\end{tikzpicture}
	\caption{Total Droplets vs. Runtime. This data was obtained by running the simulation at a constant grid size of 45 $\times$ 45 over a range of gene lengths two to twelve. The number above each point represents the gene length. The vertical axis displays the total congestion and the horizontal axis displays the runtime of the simulation. The relationship between the number of pulled droplets and runtime is approximately exponential.}
	\label{fig:RunVsTotDrops}
\end{figure}

To analyze congestion, the program was run on a range of gene lengths of two to twelve on a single machine (Machine 3 in Table \ref{t1}) at a constant grid size (45 $\times$ 45). For each gene length, the number of droplets pulled, the maximum number of concurrent droplets, and the grid's maximum congestion were recorded. Congestion is computed as the number of routed, inhabited, and occluded grid points divided by the total number of grid points. The number of droplets pulled is directly related to the length of the target gene. In Fig. \ref{fig:TotDropVsMaxDrop} and Fig. \ref{fig:TotDropVsMaxCong}, data points are labeled with their gene length for reference. As the total number of droplets increases (due to increasing gene length) the maximum number of concurrent droplets and maximum congestion increase linearly. Analyzing runtime again but in the context of congestion, we see that as the number of total droplets pulled from reservoirs increases, the runtime increases approximately exponentially (Fig. \ref{fig:RunVsTotDrops}).

\section{Conclusions} \label{sec:Conclusions}

Ever since Watson and Crick first described the molecular structure of DNA, its information-bearing potential has been apparent to computer scientists. In principle, DNA could provide a storage medium that is many orders of magnitude denser than conventional media. Spurred by the biotech and pharma industries, the technology for both sequencing (reading) and synthesizing (writing) DNA has progressed rapidly. Nevertheless, a large gap remains between what is theoretically possible in terms of reading/writing speed and what has been demonstrated in practice. The industrial partner in this research, Seagate, is developing a digital microfluidic device to close the gap. 

This paper discusses our strategy for DNA synthesis with this device and profiles the software that will control it. DNA storage units called ``genes'' are assembled from smaller DNA fragments. A key innovation is the use of a dual library of DNA fragments: ``symbols'' and ``linkers.'' Data is conveyed through the sequence of nucleotides in the symbols. Linkers, attached to the ends of symbols, determine in what order these symbols will link together. The linkers allow parallelization in the assembly process: multiple symbols can be linked together in the same droplet on the digital microfluidic device, with the linkers assuring that they hybridize in the correct order. This parallelism in synthesis is the key to achieving the write speeds needed for a DNA storage device to compete with existing systems that use magnetic, optical, or solid-state media.

The digital microfluidic device that Seagate is building will be on a scale far greater than any built with this technology today. The software to control it has to route thousands of droplets across this grid to assemble the target DNA genes. We discussed the architecture of the software that we developed for this task and presented simulation results profiling its performance. The core of the software is the routing algorithm: a prioritized 3-D A* algorithm, with two of the dimensions being the $x$-$y$ coordinates of the electrodes, and the third being time. 

Contrary to expectations, we found that runtime was limited chiefly by CPU usage; RAM usage remained manageable. When grid size was swept from 500 $\times$ 500 to 1500 $\times$ 1500, it was seen that runtime grew exponentially, RAM usage increased linearly, and CPU usage remained constant. The majority of the runtime was spent advancing node instructions and moving droplets while time spent routing droplets with the 3-D A* algorithm was relatively less.

In future work, we plan to explore routing on the device under conditions of extreme congestion, that is to say when droplets occupy nearly all the available electrodes. The simulation must be parameterized with congestion in mind; factors such as grid size, number of reaction sites, and gene length influence grid congestion and consequently runtime. It might be advantageous to incorporate algorithms designed for memory management if peak RAM becomes the limiting factor for large grid sizes. Algorithms like A* and Moving Target D* Lite are heuristic-based. They will find the shortest path under given constraints. However, they do not consider the search time and memory requirements necessary to find such a path. There exists a family of algorithms called Conflict-Based Search (CBS) which help prune unnecessary branches of the search tree in order to manage memory and improve speed. Conflict-based searching might be particularly efficient if the problem is formulated in terms of multiagent path finding ~\cite{ren2022conflict, li2022mapf, pianpak2021dmapf}.

Finally, this paper did not consider the process of reading data stored in DNA (i.e., sequencing it). With current technology, reading DNA is orders of magnitude more efficient than writing it, so the impetus is to focus on improvements in write speed. In future work, we will prototype and report experimental results on the complete system: a digital microfluidic device for writing DNA at a high speed to compete with existing solid-state storage media. 

\section*{Author Contributions}

A. Manicka is the primary author. A. Stephan designed and implemented the software and also contributed substantial portions of the text. S. Chari analyzed and improved the core algorithms for droplet routing. G. Mendonsa devised the synthesis procedure with symbols and linkers, along with colleagues at Seagate. She also contributed the background material on Gibson assembly. P. Okubo performed software testing on the system. J. Stolzberg-Schray wrote the description of the A* routing algorithm. A. Reddy and M. Riedel were the principal investigators for the project. All authors have contributed to the writing of this article.

\section*{Acknowledgements}
We would like to thank Matthew Boros and Hershen Nair for their comments and edits that greatly improved the manuscript. This work was funded in part by the National Science Foundation's Division of Computing and Communication Foundations, grant \#2227578.

\section*{Conflicts of interest}
There are no conflicts to declare.




\balance


\scriptsize{
	\bibliography{refs} 

\providecommand*{\mcitethebibliography}{\thebibliography}
\csname @ifundefined\endcsname{endmcitethebibliography}
{\let\endmcitethebibliography\endthebibliography}{}
\begin{mcitethebibliography}{60}
\providecommand*{\natexlab}[1]{#1}
\providecommand*{\mciteSetBstSublistMode}[1]{}
\providecommand*{\mciteSetBstMaxWidthForm}[2]{}
\providecommand*{\mciteBstWouldAddEndPuncttrue}
  {\def\EndOfBibitem{\unskip.}}
\providecommand*{\mciteBstWouldAddEndPunctfalse}
  {\let\EndOfBibitem\relax}
\providecommand*{\mciteSetBstMidEndSepPunct}[3]{}
\providecommand*{\mciteSetBstSublistLabelBeginEnd}[3]{}
\providecommand*{\EndOfBibitem}{}
\mciteSetBstSublistMode{f}
\mciteSetBstMaxWidthForm{subitem}
{(\emph{\alph{mcitesubitemcount}})}
\mciteSetBstSublistLabelBeginEnd{\mcitemaxwidthsubitemform\space}
{\relax}{\relax}

\bibitem[Li \emph{et~al.}(2017)Li, Stones, Wang, Liu, Li, and Xu]{li2017hard}
J.~Li, R.~J. Stones, G.~Wang, X.~Liu, Z.~Li and M.~Xu, \emph{Reliability
  Engineering \& System Safety}, 2017, \textbf{164}, 55--65\relax
\mciteBstWouldAddEndPuncttrue
\mciteSetBstMidEndSepPunct{\mcitedefaultmidpunct}
{\mcitedefaultendpunct}{\mcitedefaultseppunct}\relax
\EndOfBibitem
\bibitem[IDC(2021)]{reinsel2021worldwide}
IDC, \emph{Worldwide Global DataSphere Forecast, 2021–2025: The World Keeps
  Creating More Data — Now, What Do We Do with It All?, IDC Doc
  {\#}US46410421}, 2021\relax
\mciteBstWouldAddEndPuncttrue
\mciteSetBstMidEndSepPunct{\mcitedefaultmidpunct}
{\mcitedefaultendpunct}{\mcitedefaultseppunct}\relax
\EndOfBibitem
\bibitem[Mellor(2022)]{archiving}
C.~Mellor, \emph{Zettabyte era brings archiving front and center}, 2022,
  \url{https://blocksandfiles.com/2022/07/11/zettabyte-era-brings-archiving-front-and-center/}\relax
\mciteBstWouldAddEndPuncttrue
\mciteSetBstMidEndSepPunct{\mcitedefaultmidpunct}
{\mcitedefaultendpunct}{\mcitedefaultseppunct}\relax
\EndOfBibitem
\bibitem[IDC(March 2021)]{IDC2021storage}
IDC, \emph{Worldwide Global StorageSphere Forecast, 2021–2025: To Save or Not
  to Save Data, That Is the Question, IDC Doc {\#}US47509621}, March 2021\relax
\mciteBstWouldAddEndPuncttrue
\mciteSetBstMidEndSepPunct{\mcitedefaultmidpunct}
{\mcitedefaultendpunct}{\mcitedefaultseppunct}\relax
\EndOfBibitem
\bibitem[Monroe and Preston(July 2020)]{gartnerMarketTrends}
J.~Monroe and R.~Preston, \emph{Gartner Inc., Market Trends: Evolving
  Enterprise Data Requirements — How Much Is Not Enough?}, July 2020\relax
\mciteBstWouldAddEndPuncttrue
\mciteSetBstMidEndSepPunct{\mcitedefaultmidpunct}
{\mcitedefaultendpunct}{\mcitedefaultseppunct}\relax
\EndOfBibitem
\bibitem[Leproust(2022)]{Leproust2022Data}
E.~Leproust, \emph{Data Centers Are Unsustainable. We Need to Store Data in
  DNA.}, 2022\relax
\mciteBstWouldAddEndPuncttrue
\mciteSetBstMidEndSepPunct{\mcitedefaultmidpunct}
{\mcitedefaultendpunct}{\mcitedefaultseppunct}\relax
\EndOfBibitem
\bibitem[Church \emph{et~al.}(2012)Church, Gao, and Kosuri]{church12}
G.~Church, Y.~Gao and S.~Kosuri, \emph{Science (New York, N.Y.)}, 2012,
  \textbf{337}, 1628\relax
\mciteBstWouldAddEndPuncttrue
\mciteSetBstMidEndSepPunct{\mcitedefaultmidpunct}
{\mcitedefaultendpunct}{\mcitedefaultseppunct}\relax
\EndOfBibitem
\bibitem[sea()]{seagate.com}
\emph{Barracuda Fast SSD: Compact portable SSD with USB-C: Seagate US},
  \url{https://www.seagate.com/products/external-hard-drives/barracuda-fast-ssd/}\relax
\mciteBstWouldAddEndPuncttrue
\mciteSetBstMidEndSepPunct{\mcitedefaultmidpunct}
{\mcitedefaultendpunct}{\mcitedefaultseppunct}\relax
\EndOfBibitem
\bibitem[El-Shaikh \emph{et~al.}(2022)El-Shaikh, Welzel, Heider, and
  Seeger]{el2022high}
A.~El-Shaikh, M.~Welzel, D.~Heider and B.~Seeger, \emph{NAR genomics and
  bioinformatics}, 2022, \textbf{4}, lqab126\relax
\mciteBstWouldAddEndPuncttrue
\mciteSetBstMidEndSepPunct{\mcitedefaultmidpunct}
{\mcitedefaultendpunct}{\mcitedefaultseppunct}\relax
\EndOfBibitem
\bibitem[Ceze \emph{et~al.}(2019)Ceze, Nivala, and Strauss]{ceze19}
L.~Ceze, J.~Nivala and K.~Strauss, \emph{Nature Reviews Genetics}, 2019,
  \textbf{20}, 456--466\relax
\mciteBstWouldAddEndPuncttrue
\mciteSetBstMidEndSepPunct{\mcitedefaultmidpunct}
{\mcitedefaultendpunct}{\mcitedefaultseppunct}\relax
\EndOfBibitem
\bibitem[Watson and Crick(1953)]{watson1953structure}
J.~D. Watson and F.~H. Crick, Cold Spring Harbor symposia on quantitative
  biology, 1953, pp. 123--131\relax
\mciteBstWouldAddEndPuncttrue
\mciteSetBstMidEndSepPunct{\mcitedefaultmidpunct}
{\mcitedefaultendpunct}{\mcitedefaultseppunct}\relax
\EndOfBibitem
\bibitem[Chen \emph{et~al.}(2020)Chen, Zhu, Bošković, and Keyser]{chen20}
K.~Chen, J.~Zhu, F.~Bošković and U.~F. Keyser, \emph{Nano Letters}, 2020,
  \textbf{20}, 3754--3760\relax
\mciteBstWouldAddEndPuncttrue
\mciteSetBstMidEndSepPunct{\mcitedefaultmidpunct}
{\mcitedefaultendpunct}{\mcitedefaultseppunct}\relax
\EndOfBibitem
\bibitem[Dickinson \emph{et~al.}(2021)Dickinson, Mortuza, Clay, Piantanida,
  Green, Watson, Hayden, Andersen, Kuang, Graugnard, Zadegan, and
  Hughes]{dickinson21}
G.~D. Dickinson, G.~M. Mortuza, W.~Clay, L.~Piantanida, C.~M. Green, C.~Watson,
  E.~J. Hayden, T.~Andersen, W.~Kuang, E.~Graugnard, R.~Zadegan and W.~L.
  Hughes, \emph{Nature Communications}, 2021, \textbf{12}, 2371\relax
\mciteBstWouldAddEndPuncttrue
\mciteSetBstMidEndSepPunct{\mcitedefaultmidpunct}
{\mcitedefaultendpunct}{\mcitedefaultseppunct}\relax
\EndOfBibitem
\bibitem[Meares \emph{et~al.}(2022)Meares, Susumu, Mathur, Lee, Mass, Lee,
  Pensack, Yurke, Knowlton, Melinger,\emph{et~al.}]{meares2022synthesis}
A.~Meares, K.~Susumu, D.~Mathur, S.~H. Lee, O.~A. Mass, J.~Lee, R.~D. Pensack,
  B.~Yurke, W.~B. Knowlton, J.~S. Melinger \emph{et~al.}, \emph{ACS omega},
  2022, \textbf{7}, 11002--11016\relax
\mciteBstWouldAddEndPuncttrue
\mciteSetBstMidEndSepPunct{\mcitedefaultmidpunct}
{\mcitedefaultendpunct}{\mcitedefaultseppunct}\relax
\EndOfBibitem
\bibitem[Hepisuthar
  \emph{et~al.}(2021)Hepisuthar\emph{et~al.}]{hepisuthar2021comparative}
M.~Hepisuthar \emph{et~al.}, \emph{Turkish Journal of Computer and Mathematics
  Education (TURCOMAT)}, 2021, \textbf{12}, 3635--3641\relax
\mciteBstWouldAddEndPuncttrue
\mciteSetBstMidEndSepPunct{\mcitedefaultmidpunct}
{\mcitedefaultendpunct}{\mcitedefaultseppunct}\relax
\EndOfBibitem
\bibitem[Erlich and Zielinski(2017)]{erlich2017dna}
Y.~Erlich and D.~Zielinski, \emph{science}, 2017, \textbf{355}, 950--954\relax
\mciteBstWouldAddEndPuncttrue
\mciteSetBstMidEndSepPunct{\mcitedefaultmidpunct}
{\mcitedefaultendpunct}{\mcitedefaultseppunct}\relax
\EndOfBibitem
\bibitem[Jain \emph{et~al.}(2016)Jain, Olsen, Paten, and
  Akeson]{jain2016oxford}
M.~Jain, H.~E. Olsen, B.~Paten and M.~Akeson, \emph{Genome biology}, 2016,
  \textbf{17}, 1--11\relax
\mciteBstWouldAddEndPuncttrue
\mciteSetBstMidEndSepPunct{\mcitedefaultmidpunct}
{\mcitedefaultendpunct}{\mcitedefaultseppunct}\relax
\EndOfBibitem
\bibitem[Leproust(2022)]{twistbioscience}
E.~Leproust, \emph{Data Centers Are Unsustainable. We Need to Store Data in
  DNA.}, 2022, Rev. 6.0\relax
\mciteBstWouldAddEndPuncttrue
\mciteSetBstMidEndSepPunct{\mcitedefaultmidpunct}
{\mcitedefaultendpunct}{\mcitedefaultseppunct}\relax
\EndOfBibitem
\bibitem[Guo \emph{et~al.}(2022)Guo, Lian, Dong, Chen, and
  Huang]{guo2022survey}
W.~Guo, S.~Lian, C.~Dong, Z.~Chen and X.~Huang, \emph{ACM Transactions on
  Design Automation of Electronic Systems (TODAES)}, 2022, \textbf{27},
  1--33\relax
\mciteBstWouldAddEndPuncttrue
\mciteSetBstMidEndSepPunct{\mcitedefaultmidpunct}
{\mcitedefaultendpunct}{\mcitedefaultseppunct}\relax
\EndOfBibitem
\bibitem[De~Munter \emph{et~al.}(2020)De~Munter, Van~Parys, Bral, Ingels,
  Goetgeluk, Bonte, Pille, Billiet, Weening, Verhee,\emph{et~al.}]{de2020rapid}
S.~De~Munter, A.~Van~Parys, L.~Bral, J.~Ingels, G.~Goetgeluk, S.~Bonte,
  M.~Pille, L.~Billiet, K.~Weening, A.~Verhee \emph{et~al.},
  \emph{International journal of molecular sciences}, 2020, \textbf{21},
  883\relax
\mciteBstWouldAddEndPuncttrue
\mciteSetBstMidEndSepPunct{\mcitedefaultmidpunct}
{\mcitedefaultendpunct}{\mcitedefaultseppunct}\relax
\EndOfBibitem
\bibitem[Mackay \emph{et~al.}(2002)Mackay, Arden, and Nitsche]{mackay2002real}
I.~M. Mackay, K.~E. Arden and A.~Nitsche, \emph{Nucleic acids research}, 2002,
  \textbf{30}, 1292--1305\relax
\mciteBstWouldAddEndPuncttrue
\mciteSetBstMidEndSepPunct{\mcitedefaultmidpunct}
{\mcitedefaultendpunct}{\mcitedefaultseppunct}\relax
\EndOfBibitem
\bibitem[Firmansyah \emph{et~al.}(2016)Firmansyah, Masruroh, and
  Fahrianto]{firmansyah2016comparative}
E.~R. Firmansyah, S.~U. Masruroh and F.~Fahrianto, 2016 6th International
  Conference on Information and Communication Technology for The Muslim World
  (ICT4M), 2016, pp. 275--280\relax
\mciteBstWouldAddEndPuncttrue
\mciteSetBstMidEndSepPunct{\mcitedefaultmidpunct}
{\mcitedefaultendpunct}{\mcitedefaultseppunct}\relax
\EndOfBibitem
\bibitem[Yao \emph{et~al.}(2010)Yao, Lin, Xie, Wang, and Hung]{yao2010path}
J.~Yao, C.~Lin, X.~Xie, A.~J. Wang and C.-C. Hung, 2010 Seventh international
  conference on information technology: new generations, 2010, pp.
  1154--1158\relax
\mciteBstWouldAddEndPuncttrue
\mciteSetBstMidEndSepPunct{\mcitedefaultmidpunct}
{\mcitedefaultendpunct}{\mcitedefaultseppunct}\relax
\EndOfBibitem
\bibitem[Martins \emph{et~al.}(2022)Martins, Adekunle, Olaniyan, and
  Bolaji]{martins2022improved}
O.~O. Martins, A.~A. Adekunle, O.~M. Olaniyan and B.~O. Bolaji,
  \emph{Scientific African}, 2022, \textbf{15}, e01068\relax
\mciteBstWouldAddEndPuncttrue
\mciteSetBstMidEndSepPunct{\mcitedefaultmidpunct}
{\mcitedefaultendpunct}{\mcitedefaultseppunct}\relax
\EndOfBibitem
\bibitem[Goldman \emph{et~al.}(2013)Goldman, Bertone, Chen, Dessimoz, LeProust,
  Sipos, and Birney]{goldman2013towards}
N.~Goldman, P.~Bertone, S.~Chen, C.~Dessimoz, E.~M. LeProust, B.~Sipos and
  E.~Birney, \emph{nature}, 2013, \textbf{494}, 77--80\relax
\mciteBstWouldAddEndPuncttrue
\mciteSetBstMidEndSepPunct{\mcitedefaultmidpunct}
{\mcitedefaultendpunct}{\mcitedefaultseppunct}\relax
\EndOfBibitem
\bibitem[Grass \emph{et~al.}(2015)Grass, Heckel, Puddu, Paunescu, and
  Stark]{grass2015robust}
R.~N. Grass, R.~Heckel, M.~Puddu, D.~Paunescu and W.~J. Stark, \emph{Angewandte
  Chemie International Edition}, 2015, \textbf{54}, 2552--2555\relax
\mciteBstWouldAddEndPuncttrue
\mciteSetBstMidEndSepPunct{\mcitedefaultmidpunct}
{\mcitedefaultendpunct}{\mcitedefaultseppunct}\relax
\EndOfBibitem
\bibitem[Bornholt \emph{et~al.}(2016)Bornholt, Lopez, Carmean, Ceze, Seelig,
  and Strauss]{bornholt2016dna}
J.~Bornholt, R.~Lopez, D.~M. Carmean, L.~Ceze, G.~Seelig and K.~Strauss,
  Proceedings of the Twenty-First International Conference on Architectural
  Support for Programming Languages and Operating Systems, 2016, pp.
  637--649\relax
\mciteBstWouldAddEndPuncttrue
\mciteSetBstMidEndSepPunct{\mcitedefaultmidpunct}
{\mcitedefaultendpunct}{\mcitedefaultseppunct}\relax
\EndOfBibitem
\bibitem[Blawat \emph{et~al.}(2016)Blawat, Gaedke, Huetter, Chen, Turczyk,
  Inverso, Pruitt, and Church]{blawat2016forward}
M.~Blawat, K.~Gaedke, I.~Huetter, X.-M. Chen, B.~Turczyk, S.~Inverso, B.~W.
  Pruitt and G.~M. Church, \emph{Procedia Computer Science}, 2016, \textbf{80},
  1011--1022\relax
\mciteBstWouldAddEndPuncttrue
\mciteSetBstMidEndSepPunct{\mcitedefaultmidpunct}
{\mcitedefaultendpunct}{\mcitedefaultseppunct}\relax
\EndOfBibitem
\bibitem[Su \emph{et~al.}(2006)Su, Hwang, and Chakrabarty]{su2006droplet}
F.~Su, W.~Hwang and K.~Chakrabarty, Proceedings of the Design Automation \&
  Test in Europe Conference, 2006, pp. 1--6\relax
\mciteBstWouldAddEndPuncttrue
\mciteSetBstMidEndSepPunct{\mcitedefaultmidpunct}
{\mcitedefaultendpunct}{\mcitedefaultseppunct}\relax
\EndOfBibitem
\bibitem[Xu and Chakrabarty(2007)]{xu2007integrated}
T.~Xu and K.~Chakrabarty, proceedings of the 44th annual Design Automation
  Conference, 2007, pp. 948--953\relax
\mciteBstWouldAddEndPuncttrue
\mciteSetBstMidEndSepPunct{\mcitedefaultmidpunct}
{\mcitedefaultendpunct}{\mcitedefaultseppunct}\relax
\EndOfBibitem
\bibitem[Zhao and Chakrabarty(2012)]{zhao2012simultaneous}
Y.~Zhao and K.~Chakrabarty, \emph{IEEE Transactions on Computer-Aided Design of
  Integrated Circuits and Systems}, 2012, \textbf{31}, 242--254\relax
\mciteBstWouldAddEndPuncttrue
\mciteSetBstMidEndSepPunct{\mcitedefaultmidpunct}
{\mcitedefaultendpunct}{\mcitedefaultseppunct}\relax
\EndOfBibitem
\bibitem[Bohringer(2004)]{bohringer2004towards}
K.~Bohringer, IEEE International Conference on Robotics and Automation, 2004.
  Proceedings. ICRA'04. 2004, 2004, pp. 1468--1474\relax
\mciteBstWouldAddEndPuncttrue
\mciteSetBstMidEndSepPunct{\mcitedefaultmidpunct}
{\mcitedefaultendpunct}{\mcitedefaultseppunct}\relax
\EndOfBibitem
\bibitem[Tsung-Wei and Ho(2009)]{tsung2009fast}
H.~Tsung-Wei and T.~Ho, IEEE International Conference on Computer Design, 2009,
  pp. 445--450\relax
\mciteBstWouldAddEndPuncttrue
\mciteSetBstMidEndSepPunct{\mcitedefaultmidpunct}
{\mcitedefaultendpunct}{\mcitedefaultseppunct}\relax
\EndOfBibitem
\bibitem[Ju{\'a}rez \emph{et~al.}(2018)Ju{\'a}rez, Brizuela, and
  Mart{\'\i}nez-P{\'e}rez]{juarez2018evolutionary}
J.~Ju{\'a}rez, C.~A. Brizuela and I.~M. Mart{\'\i}nez-P{\'e}rez,
  \emph{Information Sciences}, 2018, \textbf{429}, 130--146\relax
\mciteBstWouldAddEndPuncttrue
\mciteSetBstMidEndSepPunct{\mcitedefaultmidpunct}
{\mcitedefaultendpunct}{\mcitedefaultseppunct}\relax
\EndOfBibitem
\bibitem[Liu \emph{et~al.}(2013)Liu, Chang, Liang, and Huang]{liu2013sample}
C.-H. Liu, H.-H. Chang, T.-C. Liang and J.-D. Huang, 2013 IEEE/ACM
  International Conference on Computer-Aided Design (ICCAD), 2013, pp.
  615--621\relax
\mciteBstWouldAddEndPuncttrue
\mciteSetBstMidEndSepPunct{\mcitedefaultmidpunct}
{\mcitedefaultendpunct}{\mcitedefaultseppunct}\relax
\EndOfBibitem
\bibitem[Lehotay and Cook(2015)]{lehotay2015sampling}
S.~J. Lehotay and J.~M. Cook, \emph{Journal of agricultural and food
  chemistry}, 2015, \textbf{63}, 4395--4404\relax
\mciteBstWouldAddEndPuncttrue
\mciteSetBstMidEndSepPunct{\mcitedefaultmidpunct}
{\mcitedefaultendpunct}{\mcitedefaultseppunct}\relax
\EndOfBibitem
\bibitem[Perut \emph{et~al.}(2016)Perut, Dallari, Rani, Baldini, and
  Granchi]{perut2016cell}
F.~Perut, D.~Dallari, N.~Rani, N.~Baldini and D.~Granchi, \emph{Current
  Pharmaceutical Biotechnology}, 2016, \textbf{17}, 1079--1088\relax
\mciteBstWouldAddEndPuncttrue
\mciteSetBstMidEndSepPunct{\mcitedefaultmidpunct}
{\mcitedefaultendpunct}{\mcitedefaultseppunct}\relax
\EndOfBibitem
\bibitem[Mugele and Baret(2005)]{mugele2005electrowetting}
F.~Mugele and J.-C. Baret, \emph{Journal of physics: condensed matter}, 2005,
  \textbf{17}, R705\relax
\mciteBstWouldAddEndPuncttrue
\mciteSetBstMidEndSepPunct{\mcitedefaultmidpunct}
{\mcitedefaultendpunct}{\mcitedefaultseppunct}\relax
\EndOfBibitem
\bibitem[Bender \emph{et~al.}(2016)Bender, Aijian, and
  Garrell]{bender2016digital}
B.~F. Bender, A.~P. Aijian and R.~L. Garrell, \emph{Lab on a Chip}, 2016,
  \textbf{16}, 1505--1513\relax
\mciteBstWouldAddEndPuncttrue
\mciteSetBstMidEndSepPunct{\mcitedefaultmidpunct}
{\mcitedefaultendpunct}{\mcitedefaultseppunct}\relax
\EndOfBibitem
\bibitem[Fair(2007)]{fair07}
R.~B. Fair, \emph{Microfluidics and Nanofluidics}, 2007, \textbf{3},
  245--281\relax
\mciteBstWouldAddEndPuncttrue
\mciteSetBstMidEndSepPunct{\mcitedefaultmidpunct}
{\mcitedefaultendpunct}{\mcitedefaultseppunct}\relax
\EndOfBibitem
\bibitem[Millington \emph{et~al.}(2018)Millington, Norton, Singh, Sista,
  Srinivasan, and Pamula]{millington18}
D.~Millington, S.~Norton, R.~Singh, R.~Sista, V.~Srinivasan and V.~Pamula,
  \emph{Expert review of molecular diagnostics}, 2018, \textbf{18},
  701--712\relax
\mciteBstWouldAddEndPuncttrue
\mciteSetBstMidEndSepPunct{\mcitedefaultmidpunct}
{\mcitedefaultendpunct}{\mcitedefaultseppunct}\relax
\EndOfBibitem
\bibitem[Yang and Ho(2021)]{yang21}
Y.-T. Yang and T.-Y. Ho, \emph{Front. Chem.}, 2021, \textbf{9}, 676365\relax
\mciteBstWouldAddEndPuncttrue
\mciteSetBstMidEndSepPunct{\mcitedefaultmidpunct}
{\mcitedefaultendpunct}{\mcitedefaultseppunct}\relax
\EndOfBibitem
\bibitem[Gibson \emph{et~al.}(2009)Gibson, Young, Chuang, Venter, Hutchison,
  and Smith]{gibson2009enzymatic}
D.~G. Gibson, L.~Young, R.-Y. Chuang, J.~C. Venter, C.~A. Hutchison and H.~O.
  Smith, \emph{Nature methods}, 2009, \textbf{6}, 343--345\relax
\mciteBstWouldAddEndPuncttrue
\mciteSetBstMidEndSepPunct{\mcitedefaultmidpunct}
{\mcitedefaultendpunct}{\mcitedefaultseppunct}\relax
\EndOfBibitem
\bibitem[Basova and Foret(2015)]{basova2015droplet}
E.~Y. Basova and F.~Foret, \emph{Analyst}, 2015, \textbf{140}, 22--38\relax
\mciteBstWouldAddEndPuncttrue
\mciteSetBstMidEndSepPunct{\mcitedefaultmidpunct}
{\mcitedefaultendpunct}{\mcitedefaultseppunct}\relax
\EndOfBibitem
\bibitem[Shim \emph{et~al.}(2013)Shim, Ranasinghe, Smith, Ibrahim, Hollfelder,
  Huck, Klenerman, and Abell]{shim2013ultrarapid}
J.-u. Shim, R.~T. Ranasinghe, C.~A. Smith, S.~M. Ibrahim, F.~Hollfelder, W.~T.
  Huck, D.~Klenerman and C.~Abell, \emph{ACS nano}, 2013, \textbf{7},
  5955--5964\relax
\mciteBstWouldAddEndPuncttrue
\mciteSetBstMidEndSepPunct{\mcitedefaultmidpunct}
{\mcitedefaultendpunct}{\mcitedefaultseppunct}\relax
\EndOfBibitem
\bibitem[Li \emph{et~al.}(2021)Li, Cao, Huang, Han, Wu, Sun, Zheng, Zhao, Ma,
  Jin,\emph{et~al.}]{li2021active}
D.~Li, Y.~Cao, B.~Huang, M.~Han, X.~Wu, Q.~Sun, C.~Zheng, L.~Zhao, C.~Ma,
  H.~Jin \emph{et~al.}, \emph{Langmuir}, 2021, \textbf{37}, 1297--1305\relax
\mciteBstWouldAddEndPuncttrue
\mciteSetBstMidEndSepPunct{\mcitedefaultmidpunct}
{\mcitedefaultendpunct}{\mcitedefaultseppunct}\relax
\EndOfBibitem
\bibitem[Newton \emph{et~al.}(1997)Newton, Graham, and Ellison]{newton1997pcr}
C.~R. Newton, A.~Graham and J.~S. Ellison, \emph{PcR}, BIOS Scientific
  Publishers Oxford, UK, 1997\relax
\mciteBstWouldAddEndPuncttrue
\mciteSetBstMidEndSepPunct{\mcitedefaultmidpunct}
{\mcitedefaultendpunct}{\mcitedefaultseppunct}\relax
\EndOfBibitem
\bibitem[Wang \emph{et~al.}(2019)Wang, Wu, Tan, and Yu]{wang20193}
J.~Wang, Z.~Wu, M.~Tan and J.~Yu, \emph{IEEE Transactions on Systems, Man, and
  Cybernetics: Systems}, 2019, \textbf{51}, 2904--2915\relax
\mciteBstWouldAddEndPuncttrue
\mciteSetBstMidEndSepPunct{\mcitedefaultmidpunct}
{\mcitedefaultendpunct}{\mcitedefaultseppunct}\relax
\EndOfBibitem
\bibitem[Bell(2009)]{bell2009hyperstar}
M.~G. Bell, \emph{Transportation Research Part B: Methodological}, 2009,
  \textbf{43}, 97--107\relax
\mciteBstWouldAddEndPuncttrue
\mciteSetBstMidEndSepPunct{\mcitedefaultmidpunct}
{\mcitedefaultendpunct}{\mcitedefaultseppunct}\relax
\EndOfBibitem
\bibitem[Zhang \emph{et~al.}(2020)Zhang, Wu, Dai, and He]{zhang2020novel}
Z.~Zhang, J.~Wu, J.~Dai and C.~He, \emph{IEEE Access}, 2020, \textbf{8},
  122757--122771\relax
\mciteBstWouldAddEndPuncttrue
\mciteSetBstMidEndSepPunct{\mcitedefaultmidpunct}
{\mcitedefaultendpunct}{\mcitedefaultseppunct}\relax
\EndOfBibitem
\bibitem[Huang \emph{et~al.}(2009)Huang, Lin, and Ho]{huang2009contamination}
T.-W. Huang, C.-H. Lin and T.-Y. Ho, 2009 IEEE/ACM International Conference on
  Computer-Aided Design-Digest of Technical Papers, 2009, pp. 151--156\relax
\mciteBstWouldAddEndPuncttrue
\mciteSetBstMidEndSepPunct{\mcitedefaultmidpunct}
{\mcitedefaultendpunct}{\mcitedefaultseppunct}\relax
\EndOfBibitem
\bibitem[Sun \emph{et~al.}(2010)Sun, Yeoh, and Koenig]{sun2010moving}
X.~Sun, W.~Yeoh and S.~Koenig, Proceedings of the 9th International Conference
  on Autonomous Agents and Multiagent Systems: volume 1-Volume 1, 2010, pp.
  67--74\relax
\mciteBstWouldAddEndPuncttrue
\mciteSetBstMidEndSepPunct{\mcitedefaultmidpunct}
{\mcitedefaultendpunct}{\mcitedefaultseppunct}\relax
\EndOfBibitem
\bibitem[Hua \emph{et~al.}(2010)Hua, Rouse, Eckhardt, Srinivasan, Pamula,
  Schell, Benton, Mitchell, and Pollack]{hua2010multiplexed}
Z.~Hua, J.~L. Rouse, A.~E. Eckhardt, V.~Srinivasan, V.~K. Pamula, W.~A. Schell,
  J.~L. Benton, T.~G. Mitchell and M.~G. Pollack, \emph{Analytical chemistry},
  2010, \textbf{82}, 2310--2316\relax
\mciteBstWouldAddEndPuncttrue
\mciteSetBstMidEndSepPunct{\mcitedefaultmidpunct}
{\mcitedefaultendpunct}{\mcitedefaultseppunct}\relax
\EndOfBibitem
\bibitem[Loveless \emph{et~al.}(2020)Loveless, Ott, and
  Brisk]{loveless2020performance}
T.~Loveless, J.~Ott and P.~Brisk, Proceedings of the 18th ACM/IEEE
  International Symposium on Code Generation and Optimization, 2020, pp.
  171--184\relax
\mciteBstWouldAddEndPuncttrue
\mciteSetBstMidEndSepPunct{\mcitedefaultmidpunct}
{\mcitedefaultendpunct}{\mcitedefaultseppunct}\relax
\EndOfBibitem
\bibitem[per()]{performancetest}
\emph{Passmark Support},
  \url{https://www.passmark.com/support/performancetest_faq/understanding-results.php}\relax
\mciteBstWouldAddEndPuncttrue
\mciteSetBstMidEndSepPunct{\mcitedefaultmidpunct}
{\mcitedefaultendpunct}{\mcitedefaultseppunct}\relax
\EndOfBibitem
\bibitem[Liu and Zhu(2018)]{8530894}
Y.~Liu and L.~Zhu, 2018 International Symposium on Networks, Computers and
  Communications (ISNCC), 2018, pp. 1--5\relax
\mciteBstWouldAddEndPuncttrue
\mciteSetBstMidEndSepPunct{\mcitedefaultmidpunct}
{\mcitedefaultendpunct}{\mcitedefaultseppunct}\relax
\EndOfBibitem
\bibitem[Katoch \emph{et~al.}(2021)Katoch, Chauhan, and
  Kumar]{katoch2021review}
S.~Katoch, S.~S. Chauhan and V.~Kumar, \emph{Multimedia Tools and
  Applications}, 2021, \textbf{80}, 8091--8126\relax
\mciteBstWouldAddEndPuncttrue
\mciteSetBstMidEndSepPunct{\mcitedefaultmidpunct}
{\mcitedefaultendpunct}{\mcitedefaultseppunct}\relax
\EndOfBibitem
\bibitem[Ren \emph{et~al.}(2022)Ren, Rathinam, and Choset]{ren2022conflict}
Z.~Ren, S.~Rathinam and H.~Choset, \emph{IEEE Transactions on Automation
  Science and Engineering}, 2022\relax
\mciteBstWouldAddEndPuncttrue
\mciteSetBstMidEndSepPunct{\mcitedefaultmidpunct}
{\mcitedefaultendpunct}{\mcitedefaultseppunct}\relax
\EndOfBibitem
\bibitem[Li \emph{et~al.}(2022)Li, Chen, Harabor, Stuckey, and
  Koenig]{li2022mapf}
J.~Li, Z.~Chen, D.~Harabor, P.~J. Stuckey and S.~Koenig, Proceedings of the
  AAAI Conference on Artificial Intelligence, 2022\relax
\mciteBstWouldAddEndPuncttrue
\mciteSetBstMidEndSepPunct{\mcitedefaultmidpunct}
{\mcitedefaultendpunct}{\mcitedefaultseppunct}\relax
\EndOfBibitem
\bibitem[Pianpak and Son(2021)]{pianpak2021dmapf}
P.~Pianpak and T.~C. Son, \emph{arXiv preprint arXiv:2109.08288}, 2021\relax
\mciteBstWouldAddEndPuncttrue
\mciteSetBstMidEndSepPunct{\mcitedefaultmidpunct}
{\mcitedefaultendpunct}{\mcitedefaultseppunct}\relax
\EndOfBibitem
\end{mcitethebibliography}
	\bibliographystyle{rsc} } 

\end{document}